\documentclass[preprint]{aastex}

\usepackage{graphicx}
\usepackage{amsmath}

\usepackage{natbib}

\newcommand{\rhoi}{\rho}

\newcommand{\pie}{p}

\newcommand{\vi}{v}

\newcommand{\xii}{\xi}

\newcommand{\velions}{{\bf v}}
\newcommand{\xiions}{{\boldsymbol\xi}}

\newcommand{\Ti}{T}

\newcommand{\kB}{k_B}

\newcommand{\mprot}{m_p}

\newcommand{\rhoiO}{\rho_0}

\newcommand{\rhoO}{\rho_0}
\newcommand{\pieO}{p_0}

\newcommand{\Hi}{H}

\newcommand{\TO}{T_0}

\newcommand{\rhobi}{\rho_b}

\newcommand{\rhobiO}{\rho_{b0}}

\newcommand{\zO}{z_0}

\newcommand{\vmax}{v_{max}}

\newcommand{\taud}{\tau_d}

\newcommand{\pO}{p_0}
\newcommand{\rhoe}{\rho_e}
\newcommand{\pU}{p_1}
\newcommand{\rhoU}{\rho_1}
\newcommand{\vU}{v_1}
\newcommand{\cs}{c_s}
\newcommand{\csi}{c_{si}}
\newcommand{\cse}{c_{se}}
\newcommand{\omegar}{\omega_R}
\newcommand{\omegai}{\omega_I}

\newcommand{\zini}{z}

\newcommand{\gs}{g_s}
\newcommand{\sO}{s_0}

\shorttitle{Dynamics of falling cold plasma blobs in the solar atmosphere}
\shortauthors{Oliver et al.}

\begin{document}

\title{Dynamics of coronal rain and descending plasma blobs in solar prominences: \\ I. Fully ionised case.}

\author{R. Oliver and R. Soler and J. Terradas}
\affil{Departament de F\'\i sica, Universitat de les Illes Balears, 07122 Palma de Mallorca, Spain}
\email{ramon.oliver@uib.es}

\and

\author{T. V. Zaqarashvili\altaffilmark{1} and M. L. Khodachenko\altaffilmark{2}}
\affil{Space Research Institute, Austrian Academy of Sciences, Schmiedlstrasse 6, 8042, \\ Graz, Austria}

\altaffiltext{1}{Also at Abastumani Astrophysical Observatory at Ilia State University, Cholokashvili Ave. 3/5, Tbilisi, Georgia}
\altaffiltext{2}{Also at Institute of Nuclear Physics, Moscow State University, Leninskie Gory, 119992, Moscow, Russia}

\begin{abstract}
Observations of active regions and limb prominences often show cold, dense blobs descending with an acceleration smaller than that of free fall. The dynamics of these condensations falling in the solar corona is investigated in this paper using a simple fully ionised plasma model. We find that the presence of a heavy condensation gives rise to a dynamical rearrangement of the coronal pressure that results in the formation of a large pressure gradient that opposes gravity. Eventually this pressure gradient becomes so large that the blob acceleration vanishes or even points upwards. Then, the blob descent is characterised by an initial acceleration phase followed by an essentially constant velocity phase. These two stages can be identified in published time-distance diagrams of coronal rain events. Both the duration of the first stage and the velocity attained by the blob increase for larger values of the ratio of blob to coronal density, for larger blob mass, and for smaller coronal temperature. Dense blobs are characterised by a detectable density growth (up to 60\% in our calculations) and by a steepening of the density in their lower part, that could lead to the formation of a shock. They also emit sound waves that could be detected as small intensity changes with periods of the order of 100~s and lasting between a few and about ten periods. Finally, the curvature of the falling path is only relevant when a very dense blob falls along inclined magnetic field lines.
\end{abstract}


\section{Introduction}

Coronal rain is a well-known example of cool material (with chromospheric to transition region temperature) falling down in the much hotter solar corona at heights of some tens of Mm. Coronal rain consists of cold, elongated blobs condensing at an active region loop and then falling down along its leg, with the loop magnetic field guiding the blobs descent \citep{beckers62}. The magnetic field structure has a much longer lifetime than the time taken for a condensation to fall \citep{beckers62} and a given loop often supports a few falling blobs. For example, \citet{antolin10b} tracked motions along 30 loops and detected more than one hundred descending condensations in a time span of 71 minutes. Moreover, \citet{antolin12a} used an 84-minute long observational sequence to study the trajectories of 2552 condensations along 242 different paths. These two examples also reveal that coronal rain events take place continuously in an active region.

Regarding the temperature of these condensations, coronal rain is usually observed in cold ``chromospheric'' lines, both of neutral atoms (Ly$\alpha$ and H$\alpha$) and of ionised elements (\ion{Ca}{2} and \ion{He}{2}). In addition, \citet{degroof04,degroof05} analyzed several coronal rain blobs by combining simultaneous images in H$\alpha$, \ion{He}{2} 304 \AA, and 171 \AA\ and concluded that there is no hot ``coronal'' counterpart to this falling material because the blobs only showed up in H$\alpha$ and \ion{He}{2}, but not in 171~\AA. The work by \citet{degroof05} also provides evidence of coupling between the neutral and ionised blobs (detected through the H$\alpha$ and \ion{He}{2} emission, respectively) because their kinematics was similar.

The first observational investigations of coronal rain gave falling speeds ranging from 40 to 60 km~s$^{-1}$ \citep{beckers62,bruzek69,foukal78,engvold79}, but no details about the blobs kinematics were given in these works. For this purpose a monitoring of their position or speed as a function of height or time is necessary. As soon as this kind of information was gathered, it became clear that the cool knots acceleration was smaller than that of free fall \citep{loughhead84,xu87,heinzel92}. A detailed tracking of many coronal rain condensations provides the following picture of their kinematics \citep{wiik96,degroof04,degroof05,antolin10b,antolin12a,antolin12b}: blobs display a free fall behavior only at the loop top, with velocities around 30--50~km~s$^{-1}$, whereas further down the loop leg the downward velocity becomes considerably smaller than that expected in a free fall regime and reaches 100--150~km~s$^{-1}$. Another description of the blobs kinematics is also given by \citet{zhang09}, who also obtained moderate accelerations and found that, in general, the blobs speed increased during the first few Mm of their trajectory, then remained roughly constant for $\sim 10$ Mm, and finally decreased abruptly when the blobs were between 5 and 10 Mm above the solar surface. Two types of descending blobs were found in this study: a ``fast'' and a ``slow'' group, whose average falling speeds were 72 km~s$^{-1}$ and 37 km~s$^{-1}$. Some authors have even computed the acceleration of these falling condensations and values smaller than the acceleration of gravity have always been derived: using data from a large ensemble of coronal rain blobs \citet{antolin10b}, \citet{antolin12a}, and \citet{antolin12b} obtained average accelerations of 100--150~m~s$^{-2}$, 83.5~m~s$^{-2}$, and 137~m~s$^{-2}$, respectively. Furthermore, \citet{schrijver01} analyzed a set of 45 coronal rain blobs and concluded that their acceleration was also smaller than that of free fall and that it had a very weak dependence on the loop inclination: it had an average value of 85 m~s$^{-2}$ for loops inclined less than 30$^\circ$ from the vertical direction, 70 m~s$^{-2}$ for loops with an average inclination between 30$^\circ$ and 47$^\circ$, and 71 m~s$^{-2}$ for a highly inclined loop with an angle of 72$^\circ$ with the radial direction.

This observational background has often been acquired with the help of either spectra or narrow band images. These techniques have the drawback that they only provide the line-of-sight projection of the blob velocity or its position on the plane of the sky, so that its kinematics is not well described. In a few works, however, spectra and images have been combined together to derive the full velocity vector. In the analysis of \citet{wiik96}, this was done from a combination of the loop shape reconstruction with the blobs Doppler shift and displacement in the plane of the sky. In the case of \citet{antolin12a,antolin12b}, the velocity vector of descending blobs was obtained by combining their apparent motion and their Doppler velocity. The important result here is that the kinematics of coronal rain blobs, namely their typical speeds and acceleration and the variation of velocity with height, do not change when the full velocity vector is considered.

Several attempts have been made to understand the formation and dynamics of coronal rain material. Numerical simulations by \citet[][see also references therein]{muller04,muller05} indicate that a constant loop heating that decreases exponentially with height results in a thermal instability that leads to catastrophic cooling at the loop apex \citep[see also][]{schrijver01}. A cold condensation then forms and falls down along the loop legs. Additional refinements were introduced by \citet{antolin10b} by taking into account three different effects, namely the variation of the loop cross sectional area, a loop heating concentrated at the base of the structure and that changes episodically in time (so as to simulate nanoflare heating events), and Alfv\'en wave dissipation at the loop feet. An important conclusion of this work is that the structure and dynamics of the coronal rain blobs are more sensitive to the pressure variations arising from catastrophic cooling than to gravity itself. This is in agreement with \citet{schrijver01}, who suggested that the internal pressure evolution of the loops, rather than gravity, determines the speed of condensations. \citet{fang13} conducted two-dimensional numerical simulations of a sheared coronal arcade with a constant heating localised near the chromosphere. About 4000 coronal rain condensations formed during 80 minutes, although the authors found that only one fourth of them would be detectable at the best present observational resolution. This suggests that a large number of coronal rain events can go unnoticed in current observations. Blob descending speeds obtained in this work range from a few to 60~km~s$^{-1}$ and in all cases are consistent with accelerations smaller than 180~m~s$^{-2}$. Finally, \citet{murawski11} put forward a completely different idea for the formation of coronal rain blobs. They showed that a pressure pulse at a magnetic null point can excite an entropy mode that creates a dense blob. Exhaustive information about coronal rain observations and numerical simulations leading to the formation of cold condensations in the corona can be found in \citet{antolin10b,antolin12a}.

Another type of cool descending blobs is also present in limb prominences, a radically different environment from active region coronal loops in which coronal rain is observed. One of the first detailed reports of this phenomenon is that of \citet{engvold76}, who used time series of H$\alpha$ filtergrams of 24 prominences located away from sunspots. He reported on bright knots that frequently appeared stationary for 2--10 min at the location where they were first detected and later descended with a speed in the plane of the sky between 15 and 35 km~s$^{-1}$. These events seem to be different from the thin vertical prominence threads in which material flows down at small speeds, of the order of 10 km~s$^{-1}$ \citep{engvold81,berger08,berger10}.

Instruments onboard Hinode and SDO have been recently used to provide detailed information about bright descending knots in prominences. With the help of H$\alpha$ and \ion{Ca}{2} H images obtained with SOT on Hinode, \citet{chae10} tracked moving emission features in a large hedgerow prominence. An important result of this work is that blobs detectable in H$\alpha$ images could always be detected in their \ion{Ca}{2} H counterparts, their descending motions showing no perceptible difference. As \citet{chae10} pointed out, this implies that \ion{Ca}{2} and neutral H are dynamically coupled. Regarding the knots kinematics, this author found that their paths are almost vertical and measured speeds in the plane-of-the-sky between 10 and 30 km~s$^{-1}$, with an acceleration that is not constant in time, although it is always below 100 m~s$^{-2}$. Plasma knots ejected in a limb prominence and then undergoing ballistic motion were studied by \citet{hillier11}. The kinematics of the descending motions were not investigated in the same detail as in \citet{chae10}, but analogous results were obtained: cool blobs had downward velocities smaller than 60 km~s$^{-1}$ and their acceleration did not remain constant during their motion.

\citet{liu12} described the formation of a prominence by the condensation of coronal mass and concluded that the prominence under study was not a static object and that at a given time during the formation process, its total mass came from the balance between mass condensation and drainage through vertical downflows. Most of the condensed mass drained down along vertical paths, the mass drainage rate being about 96\% the condensation rate. These downflows occured in the form of cool mass blobs descending with a roughly constant acceleration that is much smaller than that of free fall. A total of 874 downflowing trajectories were studied by \citet{liu12}, who concluded that a typical downflowing event lasted between a few minutes and half an hour. The descending mass blob started at a height between 20 Mm and 40 Mm and traveled a distance in the range 10--30 Mm. The derived accelerations have a mean value of 46 m~s$^{-2}$ and cover a wide range from 10 to 200~m~s$^{-2}$. Furthermore, \citet{liu12} measured the speed of blobs 25'' above the surface and values around 30 km s$^{-1}$ were found.

Explaining vertical motions in prominences is a challenging task. Most observational determinations of the magnetic field vector in quiescent prominences support the idea that it is roughly horizontal inside these objects; see, for example, \citet{leroy83,leroy84} and also the reviews by \citet{lopezariste07} and \citet{mackay10}. And because very few observations \citep[such as that by][]{merenda06} contradict this conventional belief, most theoretical efforts to model the prominence internal magnetic configuration are based on horizontal or dipped magnetic fields. Under the conditions of the frozen-free theorem, ions moving vertically must deform the roughly horizontal magnetic field lines, which means the generation of electric currents and the creation of a Lorentz force to compensate the action of ions; the situation is different for neutrals, that can slip across the magnetic field. The different dynamics of ions and neutrals in a steady-state prominence was investigated by \citet{gilbert02}, who derived falling velocities of the order of 4 m s$^{-1}$ and 80 m s$^{-1}$ for neutral H and He, respectively. The vertical drainage of neutrals (especially He) was proposed as the cause of the He deficit in the upper parts of several filaments \citep{gilbert02,gilbert07}. The descending speeds obtained in these steady-state calculations are thus much smaller than those of falling cool blobs in prominences.

\citet{haerendel11} made order-of-magnitude calculations to understand the motions of dense blobs and vertical threads in prominences. They proposed that dense plasma elements condense at the top of prominences and then sink into the mainly horizontal magnetic field. These authors postulated that the plasma next becomes squeezed vertically by the action of gravity and so a current sheet forms just above them, where magnetic field reconnection subsequently takes place. The consequence of this process is that the plasma element becomes detached from the prominence magnetic field and quickly accelerates under the action of gravity. As the plasma knot makes its way through the ambient magnetic field, it looses momentum and energy by the emission of Alfv\'en waves, and so a more or less steady vertical velocity is attained. Magnetic reconnection has also been invoked as an essential process in falling prominence condensations by \cite{chae10}. A deeper insight into this phenomenon was obtained by \citet{low12a}, who represented the prominence as a Kippenhahn-Schl\"utter plasma slab in thermal balance between optically thin radiation, heating, and conduction. \citet{low12a} suggested that the frozen-in condition may recurrently break down and so a downward resistive flow across the local horizontal magnetic field is produced, which would be detected as cold falling material in a prominence. \citet{low12a} predicted that these prominence structures may have cores much denser and less ionised than usually considered. Additional support for this hypothesis was provided by \citet{low12b}, who argued that high electrical conductivity leads to the spontaneous formation and resistive dissipation of discrete currents, and that this may lead to the dynamical situation observed in the interior of some quiescent prominences.

In this series of papers we use a very simple model corona to address several issues associated to cool descending knots both in active region loops and in quiescent prominences. First of all, we want to understand the interplay between the various forces (especially gravity and the pressure gradient, but also the inertial term) in the dynamics of these features. We deliberately ignore the formation process of these condensations and so discard thermal effects in our numerical simulations. Second, we try to understand what conditions lead to falling blobs with either constant velocity or constant acceleration. Two scenarios are possible: one in which a blob with zero initial speed is left free under the action of the various forces, and a second one in which a continuous plasma injection takes place at a certain height in the corona so that a mass condensation forms and starts to fall. While the first situation is similar to most coronal rain and descending prominence knot events, the second one resembles the observations of \citet{liu12} with a continuous condensation and downward plasma drainage. Finally, we want to assess the strength of the coupling through friction between neutrals and ions in a descending blob. Although these two components can in principle move independently \citep[as in the steady-state model of][]{gilbert02}, observations made simultaneously in suitable spectral lines show that their kinematics are the same \citep{degroof05,chae10}.

This paper is organised as follows: in \S~\ref{sect_model} the equations that describe the motions of dense, falling plasma blobs are presented. Here we concentrate in the fully ionised plasma case, while the dynamics of a partially ionised plasma blob are left for future work. In addition, only the first scenario (dynamics of a fully formed blob) is studied. \S~\ref{sect_results} and \ref{sect_fip_curvature} respectively contain the results for a vertical and a circular descending path, and conclusions are drawn in \S~\ref{sect_conclusions}.

\section{Model}
\label{sect_model}

\subsection{Governing equations}

To study the temporal evolution of a fully ionised blob falling under the influence of gravity and pressure forces, only the dynamics in the vertical direction is taken into account. Hence, the horizontal structuring is ignored to facilitate our study. Moreover, the effect of magnetic fields is ignored. This corresponds either to a non-existent magnetic field or to motions along vertical magnetic field lines. The case of low-$\beta$, high electric conductivity plasma blobs falling along curved magnetic field lines is discussed in \S~\ref{sect_fip_curvature}; the plasma $\beta$ is the ratio of the plasma pressure to the magnetic pressure.

A Cartesian coordinate system with the $z$-axis pointing in the vertical direction is considered. In addition, the plasma is assumed to be fully ionised hydrogen with a density ($\rhoi$) that is approximately that of protons and a pressure ($\pie$) equal to that of protons plus electrons. The velocity of protons is denoted by $\velions$ and it is assumed that there are only vertical mass motions, that is, $\velions=\vi {\bf \hat e}_z$. The expressions describing the temporal evolution of these variables ($\rhoi$, $\pie$, $\velions$) are the mass balance, momentum, and energy equations,

\begin{equation}\label{eqmassi}
\frac{\partial\rhoi}{\partial t}+\nabla\cdot(\rhoi\velions)=0,
\end{equation}

\begin{equation}\label{eqmomi}
\rhoi\left[\frac{\partial\velions}{\partial t}+(\velions\cdot\nabla)\velions\right]=-\nabla\pie+\rhoi{\bf g},
\end{equation}

\begin{equation}\label{eqpressie}
\frac{\partial\pie}{\partial t}+(\velions\cdot\nabla)\pie+\gamma\pie\nabla\cdot\velions=0,
\end{equation}

\noindent with ${\bf g}$ the acceleration of gravity. Joule heating and the divergence of the heat flux in the energy equation have also been neglected. All other terms missing in the energy equation vanish in the present configuration.

We supplement the above system of equations with the expression for the Lagrangian displacement of a plasma element, denoted by $\xiions$,

\begin{equation}\label{eqdispli}
\frac{\partial\xiions}{\partial t}=\velions-\velions\cdot\nabla\xiions.
\end{equation}

\noindent Thus, if a particle is initially at position ${\bf r}$, at time $t$ its position is ${\bf r}+\xiions$.

Now, we assume that variables depend on $z$ and $t$ only. Next we insert $\rhoi=\rhoi(z,t)$, $\pie=\pie(z,t)$, $\velions=\vi(z,t) {\bf \hat e}_z$, and $\xiions=\xii(z,t) {\bf \hat e}_z$ into Equations~(\ref{eqmassi})--(\ref{eqdispli}) and substitute ${\bf g}=-g{\bf \hat e}_z$. The horizontal components of the momentum equation and of the expression for the Lagrangian displacement are identically zero and so we end up with four non-linear partial differential equations for the four unknowns,

\begin{equation}\label{eqrhoi}
\frac{\partial\rhoi}{\partial t}=-\vi\frac{\partial\rhoi}{\partial z}-\rhoi\frac{\partial\vi}{\partial z},
\end{equation}

\begin{equation}\label{eqvi}
\rhoi\frac{\partial\vi}{\partial t}=-\rhoi\vi\frac{\partial\vi}{\partial z}-\frac{\partial\pie}{\partial z}-g\rhoi,
\end{equation}

\begin{equation}\label{eqpie}
\frac{\partial\pie}{\partial t}=-\vi\frac{\partial\pie}{\partial z}-\gamma\pie\frac{\partial\vi}{\partial z},
\end{equation}

\begin{equation}\label{eqxii}
\frac{\partial\xii}{\partial t}=\vi\left(1-\frac{\partial\xii}{\partial z}\right).
\end{equation}

These expressions are supplemented with the ideal gas law for a fully ionised plasma,

\begin{equation}\label{eq_temp}
\pie=2\rhoi R\Ti,
\end{equation}

\noindent with $R=\kB/\mprot$ the ideal gas constant; $\kB$ and $\mprot$ are Boltzmann's constant and the proton mass. The reason for the presence of the factor 2 in Equation~(\ref{eq_temp}) is that the partial pressures of ions and electrons are summed up under the conditions of same temperature and same number density.

\subsection{Static equilibrium}

Ignoring temporal variations and assuming $\vi=0$ in Equations~(\ref{eqrhoi})--(\ref{eqpie}) we have the equation

\begin{equation}\label{equilin}
0=-\frac{\partial\pie}{\partial z}-g\rhoi.
\end{equation}

\noindent Here we adopt the following exponentially stratified solution to Equation~(\ref{equilin})

\begin{equation}\label{sol_equili}
\pie(z,t=0)=\pieO e^{-z/\Hi}, \hspace{1cm} \rhoi(z,t=0)=\rhoiO e^{-z/\Hi},
\end{equation}

\noindent where the density and pressure at the coronal base ($z=0$) satisfy the ideal gas law,

\begin{equation}\label{eq_pieO_pnO}
\pieO=2\rhoiO R\TO,
\end{equation}

\noindent with $\TO$ the equilibrium temperature, assumed uniform. Furthermore, the vertical scale height ($\Hi$) depends on the equilibrium temperature as follows,

\begin{equation}\label{eq_Hi_Hn}
\Hi=2\frac{R\TO}{g}.
\end{equation}

\subsection{Mass condensation}
\label{sect_mass_condensation}

A fully ionised dense blob is superimposed to the previous static atmosphere and its temporal evolution is then investigated. Since the condensation process is ignored, a fully formed blob is initially at rest at a certain height and is left to evolve under the action of gravity, the pressure gradient, and the inertial term according to Equations~(\ref{eqrhoi})--(\ref{eqxii}). At $t=0$ the blob has a density described by

\begin{equation}\label{rhoblob}
\rhobi(z,t=0)=\rhobiO\exp\left[-\left(\frac{z-\zO}{\Delta}\right)^2\right],
\end{equation}

\noindent where $\rhobiO$ is the central mass density of the condensation, $\zO$ is its initial position, and its length in the vertical direction is of the order of $2\Delta$.

Then, the initial density distribution is the sum of $\rhoi$ in Equation~(\ref{sol_equili}) and $\rhobi$ in Equation~(\ref{rhoblob}). On the other hand, the initial pressure is still described by $\pie$ in Equation~(\ref{sol_equili}), so that at $t=0$ the blob is not in mechanical equilibrium and starts to fall.

\section{Numerical procedure}
\label{sect_numerical_procedure}

To summarise the expressions of \S~\ref{sect_model}, the temporal evolution of the system is described by Equations~(\ref{eqrhoi}) to (\ref{eqxii}), where the unknowns are the density, vertical velocity component, pressure, and Lagrangian displacement ($\rhoi$, $\vi$, $\pie$, and $\xii$). The initial state is given by Equations~(\ref{sol_equili}) and (\ref{rhoblob}), with the velocity and the plasma displacement both initially set to zero ($\vi=\xii=0$). We focus our attention on the vertical dynamics of descending matter in the range of heights $0\leq z\leq 60$ Mm, with the blob released at an initial height $\zO=50$~Mm. As we show later, inserting the dense condensation in an otherwise equilibrium corona results in the generation of two sound waves that propagate up and down at the coronal sound speed from the blob initial position. To avoid the influence of wave reflections coming from the boundaries, these are placed well away from our region of interest. Thus, the numerical integration is carried out in the range $-200\,\,{\rm Mm}\leq z\leq 250$ Mm. The numerical mesh consists of three regions with uniform spacing: in the ranges $-200\,\,{\rm Mm}\leq z\leq -40$ Mm and $60\,\,{\rm Mm}\leq z\leq 250$ Mm the grid spacing is 100 km, whereas in the range $-40\,\,{\rm Mm}\leq z\leq 60$ Mm it is 10 km. The PDE2D code \citep{sewell05} with Galerkin's method and linear finite elements is used for the numerical integration. Finally, during the whole simulation $\rhoi$, $\pie$, and $\xii$ at the bottom and top boundaries are fixed at their initial values, while the $z$-derivative of $\vi$ is set equal to zero there. This choice of boundary conditions provides satisfactory numerical stability.

\section{Results}
\label{sect_results}

In \S~\ref{sect_fip_blob_dynamics} we consider a set of  ``reference'' parameter values to study the blob dynamics with a numerical simulation lasting 1000~s. In the initial state the coronal gas has temperature $\TO=2\times10^6$~K (so that the vertical scale height is $\Hi\simeq 120$ Mm) and density at the coronal base $\rhoiO=5\times 10^{-12}$~kg~m$^{-3}$ (corresponding to a particle density $\simeq 3\times 10^9$ cm$^{-3}$). The base pressure is $0.165$ Pa. The initial mass condensation density is $\rhobiO=10^{-10}$~kg~m$^{-3}$, its initial position is $\zO=50$~Mm, and the parameter $\Delta$, that determines the blob length, is 0.5~Mm. In \S~\ref{sect_fip_blob_dynamics10} we present the results of a similar numerical simulation with a blob ten times denser ($\rhobiO=10^{-9}$~kg~m$^{-3}$) and with all other parameter values unchanged.

\subsection{Blob dynamics}
\label{sect_fip_blob_dynamics}

\begin{figure}[h]
  \centerline{
    \scriptsize{(a)}
    \includegraphics[width=0.31\textwidth,angle=-90]{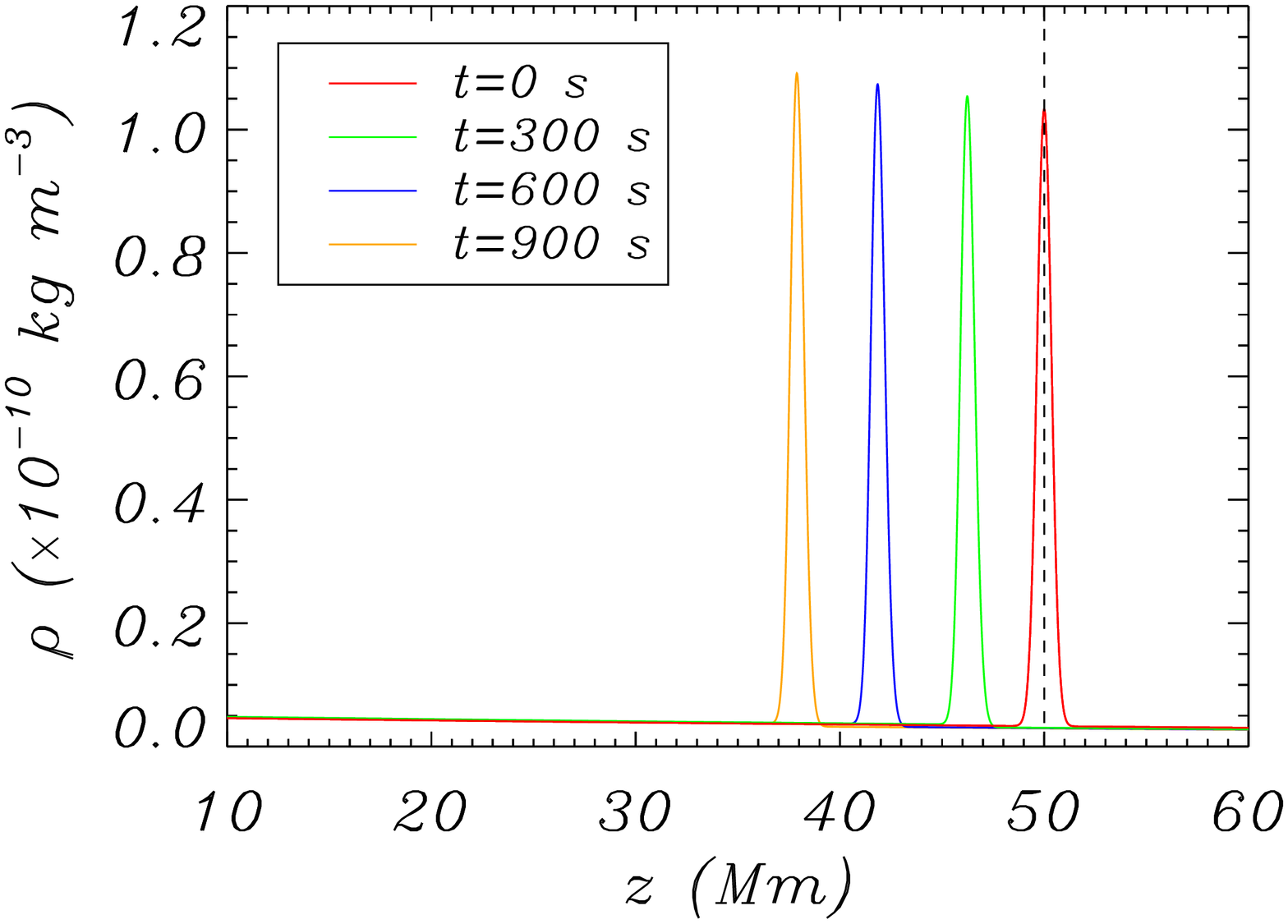} \\
    \scriptsize{(b)}
    \includegraphics[width=0.31\textwidth,angle=-90]{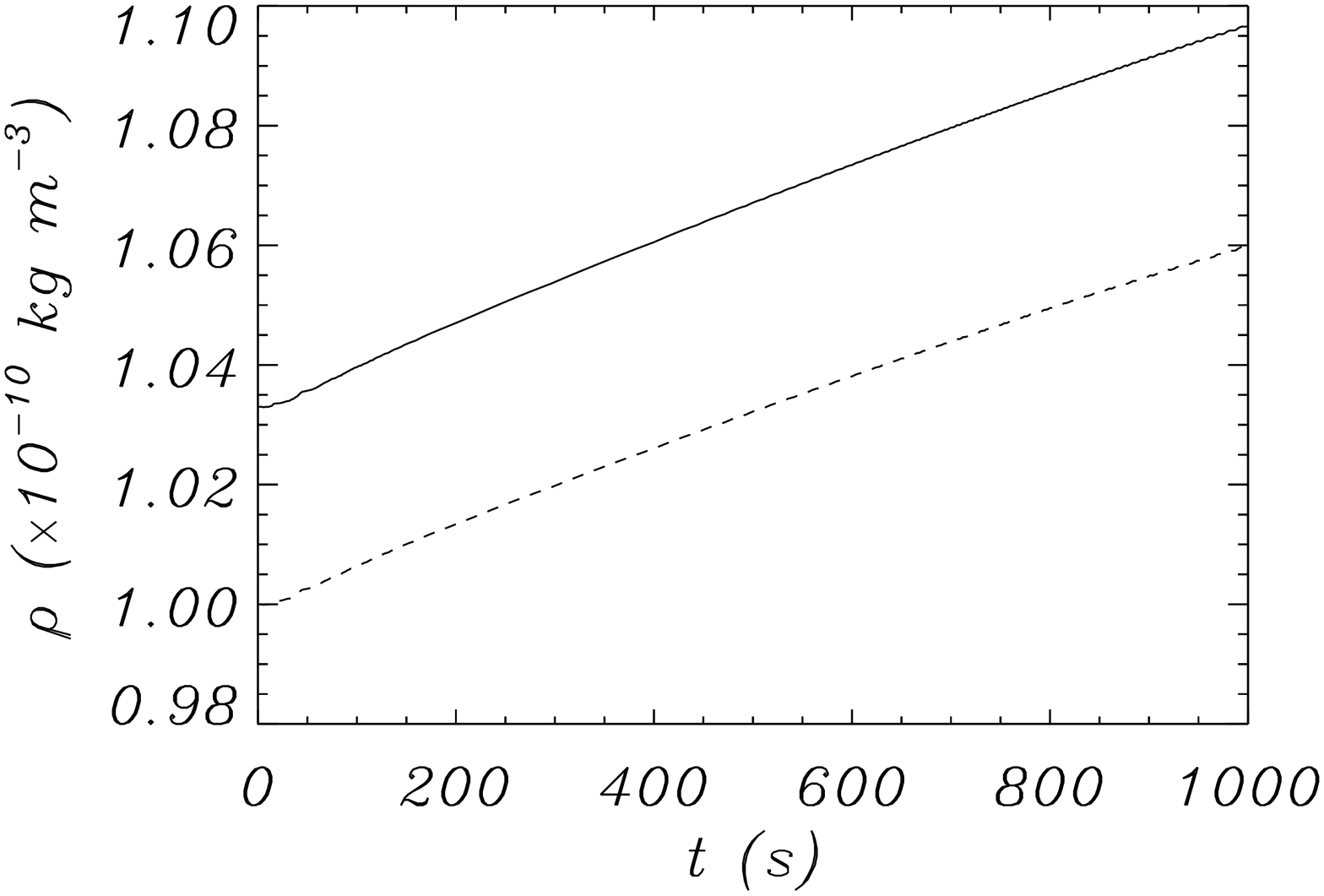} \\
  }
  \centerline{
    \scriptsize{(c)}
    \includegraphics[width=0.31\textwidth,angle=-90]{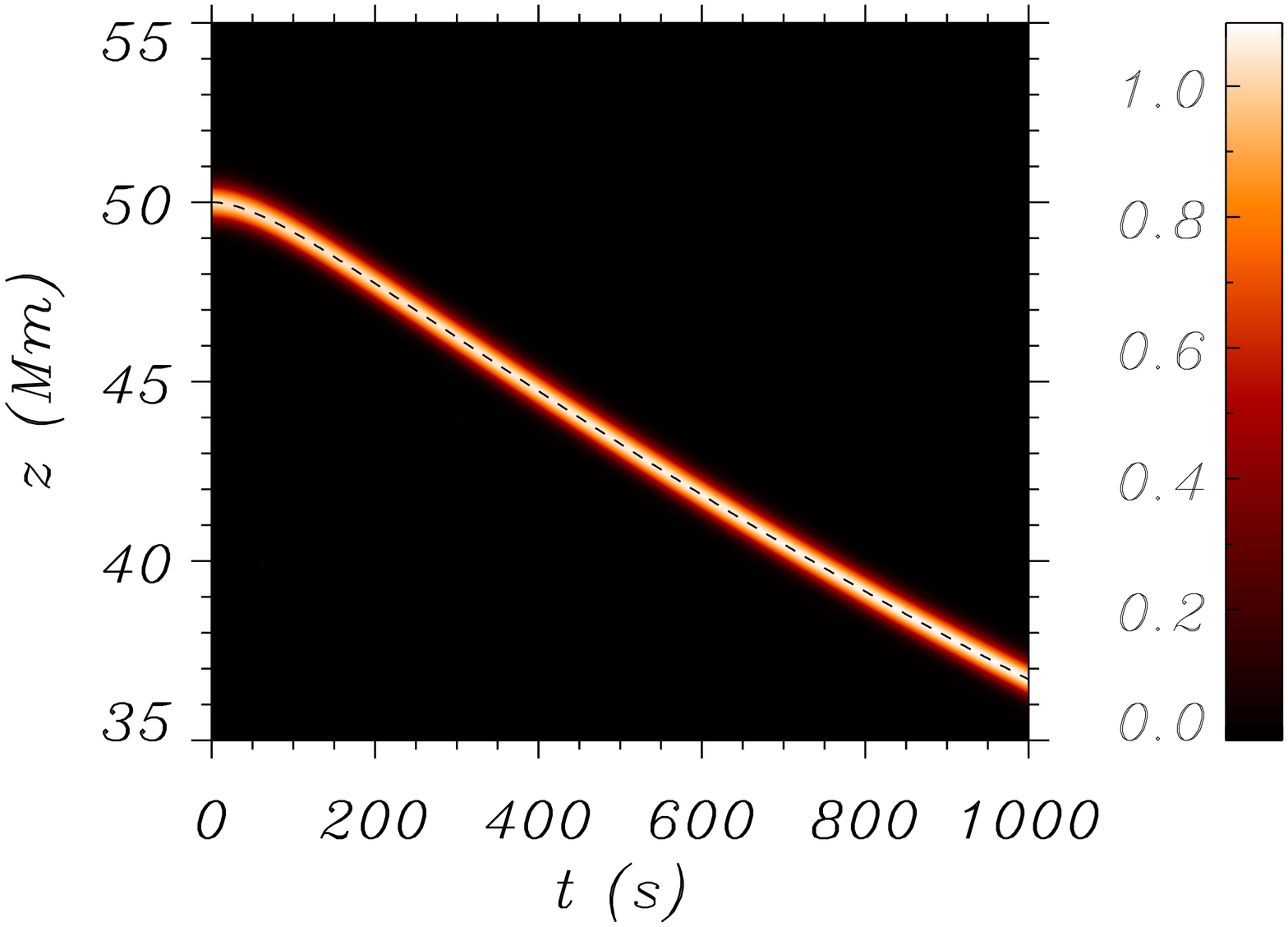}
    \scriptsize{(d)}
    \includegraphics[width=0.31\textwidth,angle=-90]{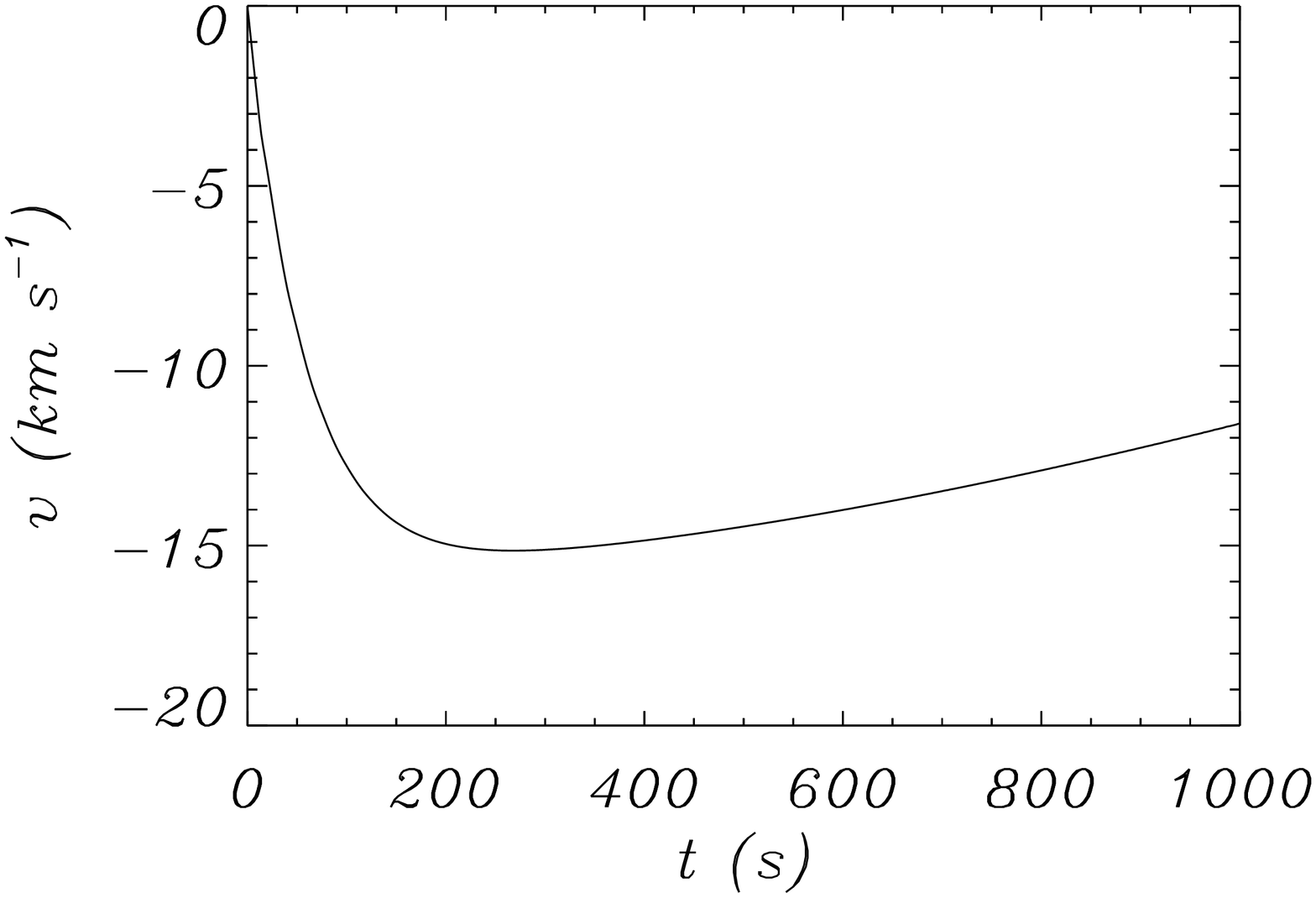}
  }
  \caption{(a) Vertical distribution of the plasma density for several times. The vertical line denotes the initial blob position. (b) Solid line: maximum blob density as a function of time. Dashed line: maximum blob density with the background coronal density subtracted. (c) Time-height diagram of the plasma density in units of $10^{-10}$~kg~m$^{-3}$. The dashed line gives the height of maximum density as a function of time. (d) Plasma velocity at the maximum density position as a function of time. (c) and (d) give the evolution of the blob height and speed versus time.Ê(An animation of panel (a) is available in the online journal.)}
  \label{fip_density}
\end{figure}

Figure~\ref{fip_density}a presents several snapshots of the temporal evolution of the density. The mass blob starts to fall under the action of gravity and apparently maintains its shape during its descending motion, although its maximum density increases slightly. From Figure~\ref{fip_density}a it is straightforward to appreciate that the blob does not fall with the acceleration of gravity: if this was the case, then the blob would reach $z=0$ at $t\simeq 540$~s and it would be outside the horizontal range of this figure. Next we draw our attention to the variation of the blob peak density in Figure~\ref{fip_density}a, which could be caused by the fact that, as the condensation falls, it moves in a denser environment. To check this possibility, in Figure~\ref{fip_density}b we represent the maximum blob density versus time (solid line) and the same quantity with the coronal density subtracted (dashed line). Both quantities display the same temporal increase and so the enhancement of the blob density is a real effect not caused by the change in the background coronal density with height. This issue is later discussed in more detail. The change of blob height with time is presented in Figure~\ref{fip_density}c. The blob trajectory is convex in the first $\simeq 200$~s and then becomes roughly linear for $t\geq 200$~s. Hence, the mass condensation accelerates downwards in the initial 200~s but then its acceleration seems to become practically zero and so the blob appears to fall with a constant velocity. This behavior can be found in time-distance diagrams of coronal rain events, as for example Figure~6 of \citet{degroof04}, Figure~5 of \citet{degroof05}, Figure~3 of \citet{antolin10b}, and Figure~7 of \citet{antolin12a}, and of descending prominence knots during the constant velocity phase \citep[see Figures~3--5 of][]{chae10}.

\begin{figure}[h]
  \centerline{
    \scriptsize{(a)}
    \includegraphics[width=0.33\textwidth,angle=-90]{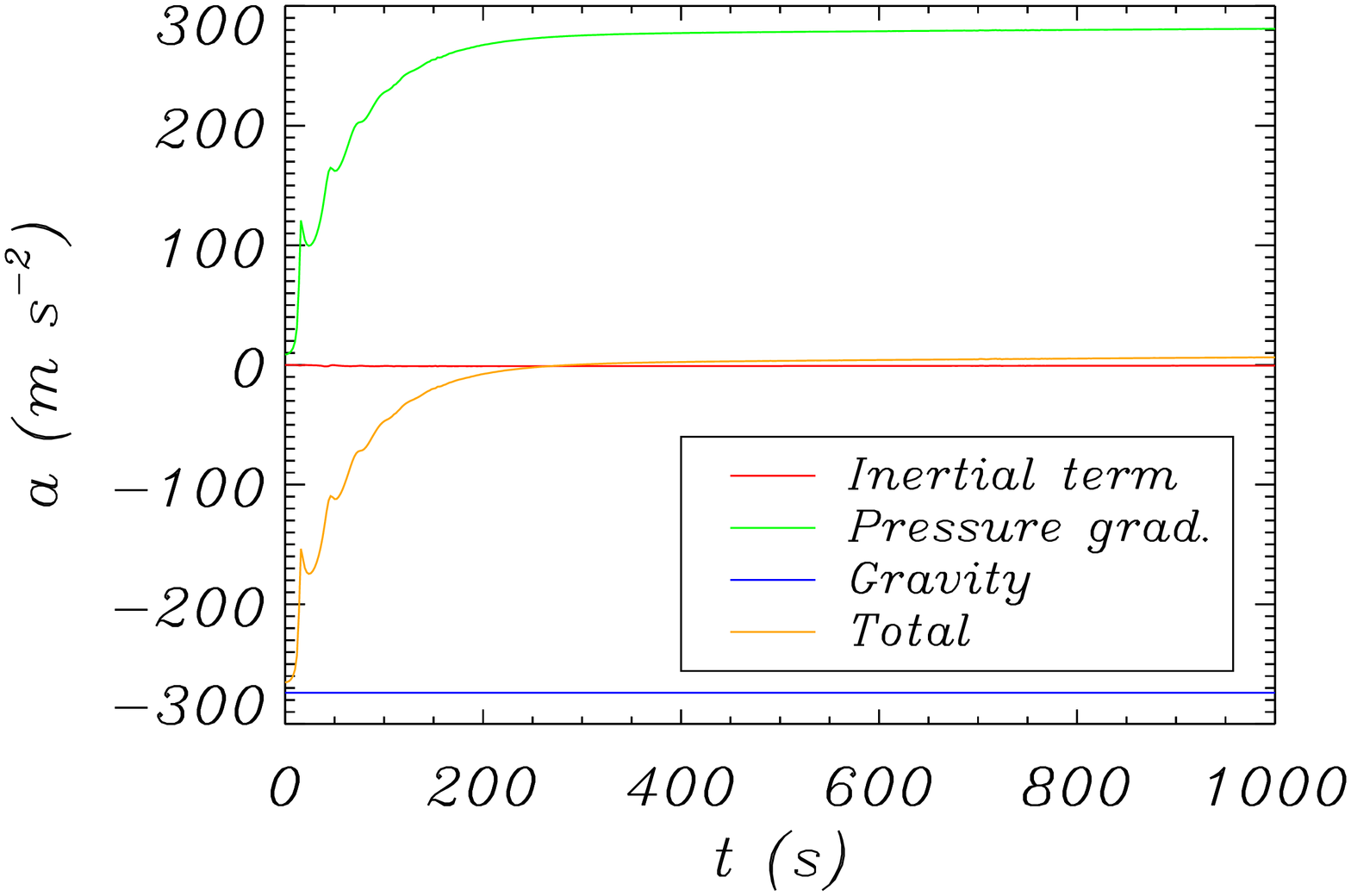}
    \scriptsize{(b)}
    \includegraphics[width=0.33\textwidth,angle=-90]{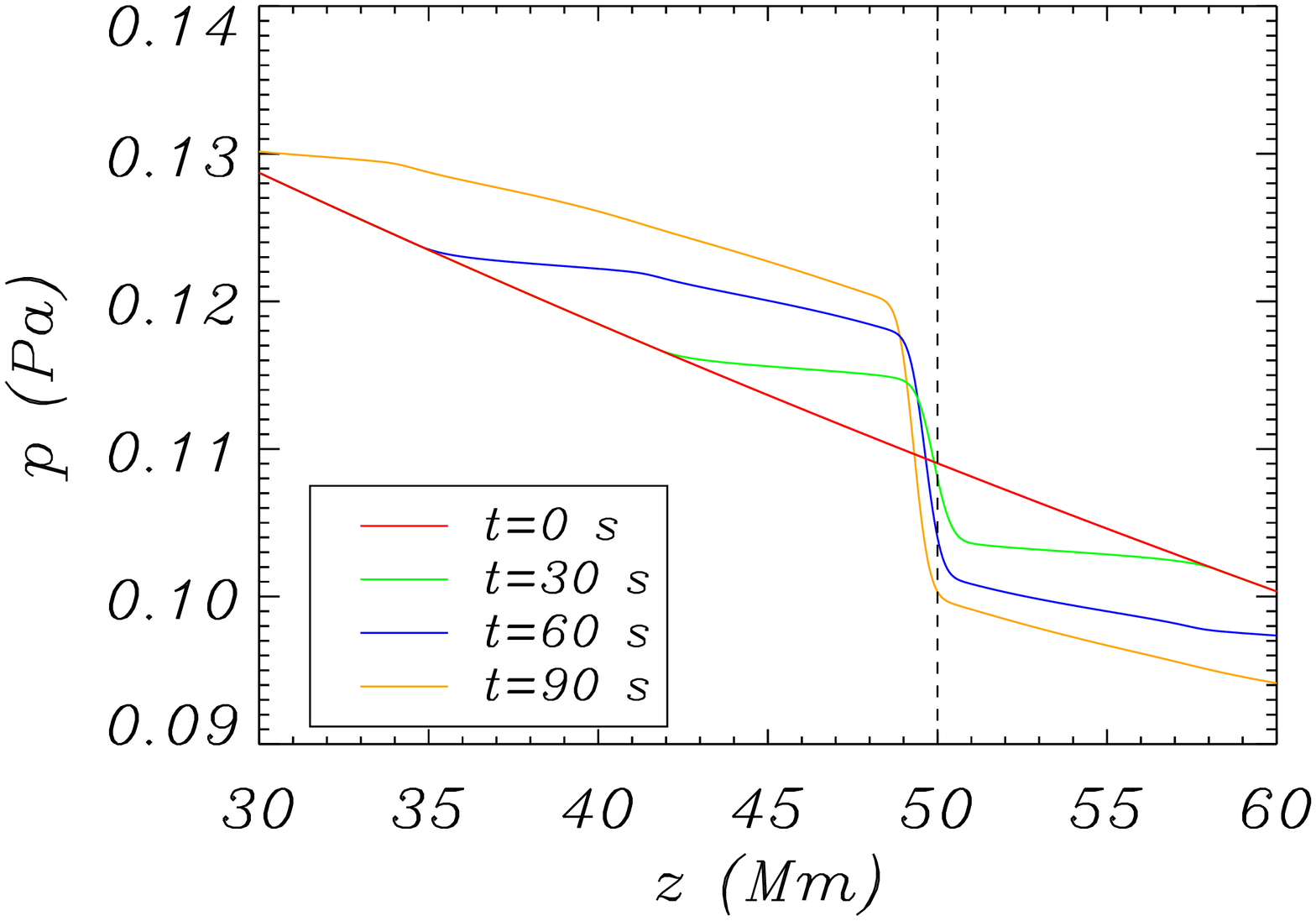}
  }
  \caption{(a) Blob acceleration caused by the terms in the momentum equation as a function of time. (b) Vertical distribution of the plasma pressure for several times. The vertical dashed line marks the initial blob position.Ê(An animation of panel (b) is available in the online journal. In this animation, the red curve is the unperturbed initial pressure and the green line shows the maximum density position.)}
  \label{fip_dynamics}
\end{figure}

To understand these two phases in the blob kinematics, we consider the path of the blob maximum density position, i.e., the pairs of values $(z,t)$ of the dashed line in Figure~\ref{fip_density}c. The blob falling speed is just the ions velocity obtained in the numerical simulation for these specific $z$ and $t$; this quantity is shown in Figure~\ref{fip_density}d. One can appreciate a rapid velocity increase (in absolute value) in the first 250~s followed by a slow deceleration that lasts for the rest of the simulation: these are the two phases mentioned before. The blob descending speed in the first stage reaches the maximum (unsigned) value $\vmax\simeq15$~km~s$^{-1}$, while in the second stage the blob speed decreases from 15~km~s$^{-1}$ to 12~km~s$^{-1}$ and so it is essentially constant although with a small upward acceleration.

Next, Equation~(\ref{eqvi}) is divided by $\rhoi$ and so the terms on the right-hand side correspond to the acceleration caused by the inertial term, the pressure gradient, and gravity, respectively. These three accelerations, together with their sum, are plotted at the  position of maximum blob density in Figure~\ref{fip_dynamics}a. We see that the inertial term causes a negligible acceleration and that the pressure gradient results in no acceleration at $t=0$. Nevertheless, the pressure gradient force increases in time and counteracts the blob weight at $t\simeq 250$~s. After this time the pressure gradient force overcomes the blob weight and the blob acceleration becomes slightly positive, as anticipated from the blob velocity in Figure~\ref{fip_density}d.

We now turn our attention to the plasma pressure and plot it as a function of height for various times at the beginning of the numerical simulation (Figure~\ref{fip_dynamics}b). The presence of a mass condensation out of mechanical equilibrium in the corona generates a sound wave that propagates up and down at the (constant) sound speed, that in a fully ionised plasma at $2\times 10^6$~K is around 235~km~s$^{-1}$. This sound wave produces a significant modification of the pressure structure such that a steep pressure gradient builds up at the blob position. As the blob travels down, so does this pressure gradient, such as can be seen in the online movie accompanying Figure~\ref{fip_dynamics}b. Then, we conclude that the vertical acceleration provided by the pressure gradient comes from a dynamical rearrangement of the plasma pressure in the corona.

\subsection{Blob dynamics for higher blob density}
\label{sect_fip_blob_dynamics10}

\begin{figure}[h]
  \centerline{
    \scriptsize{(a)}
    \includegraphics[width=0.33\textwidth,angle=-90]{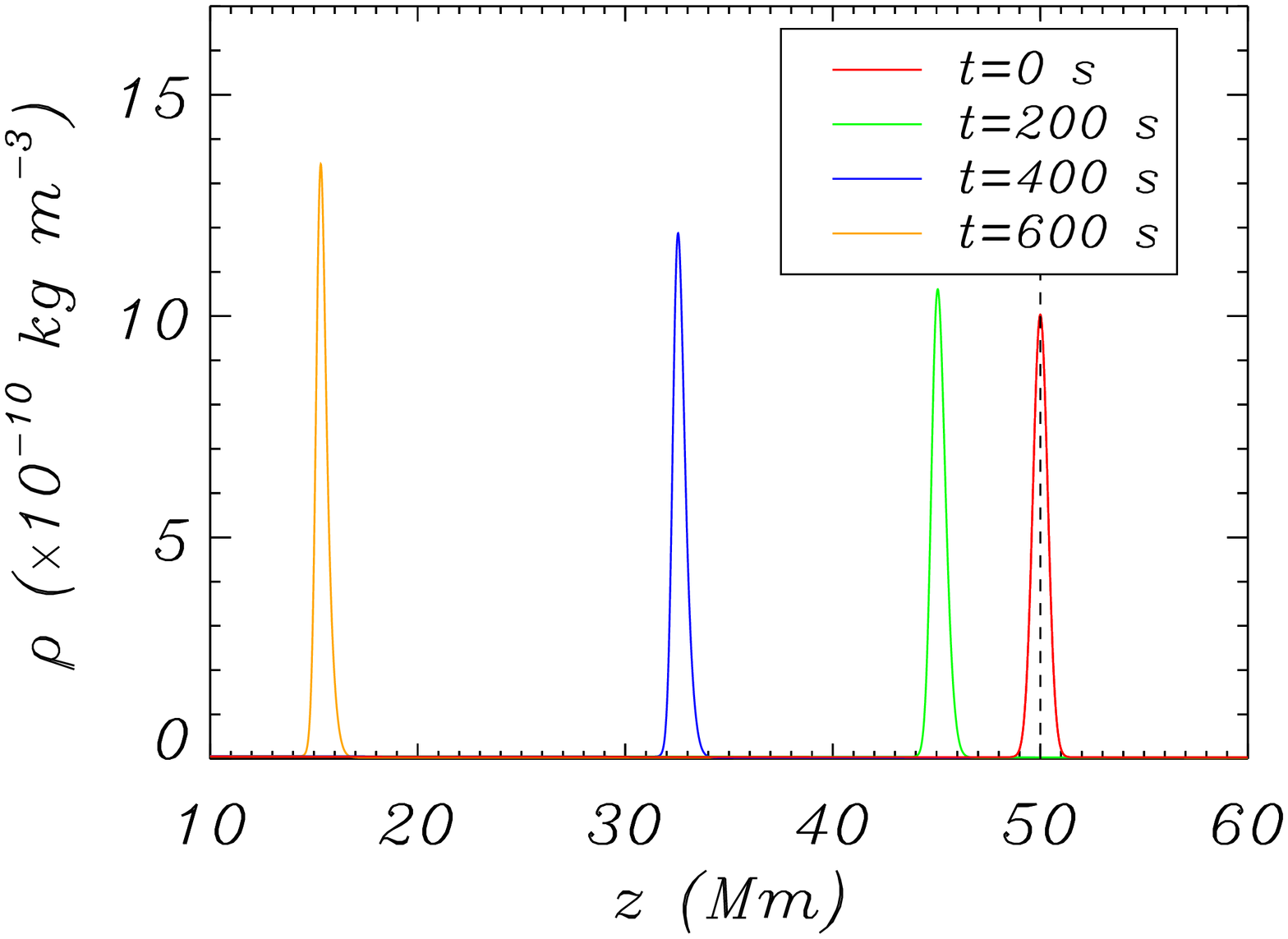} \\
    \scriptsize{(b)}
    \includegraphics[width=0.33\textwidth,angle=-90]{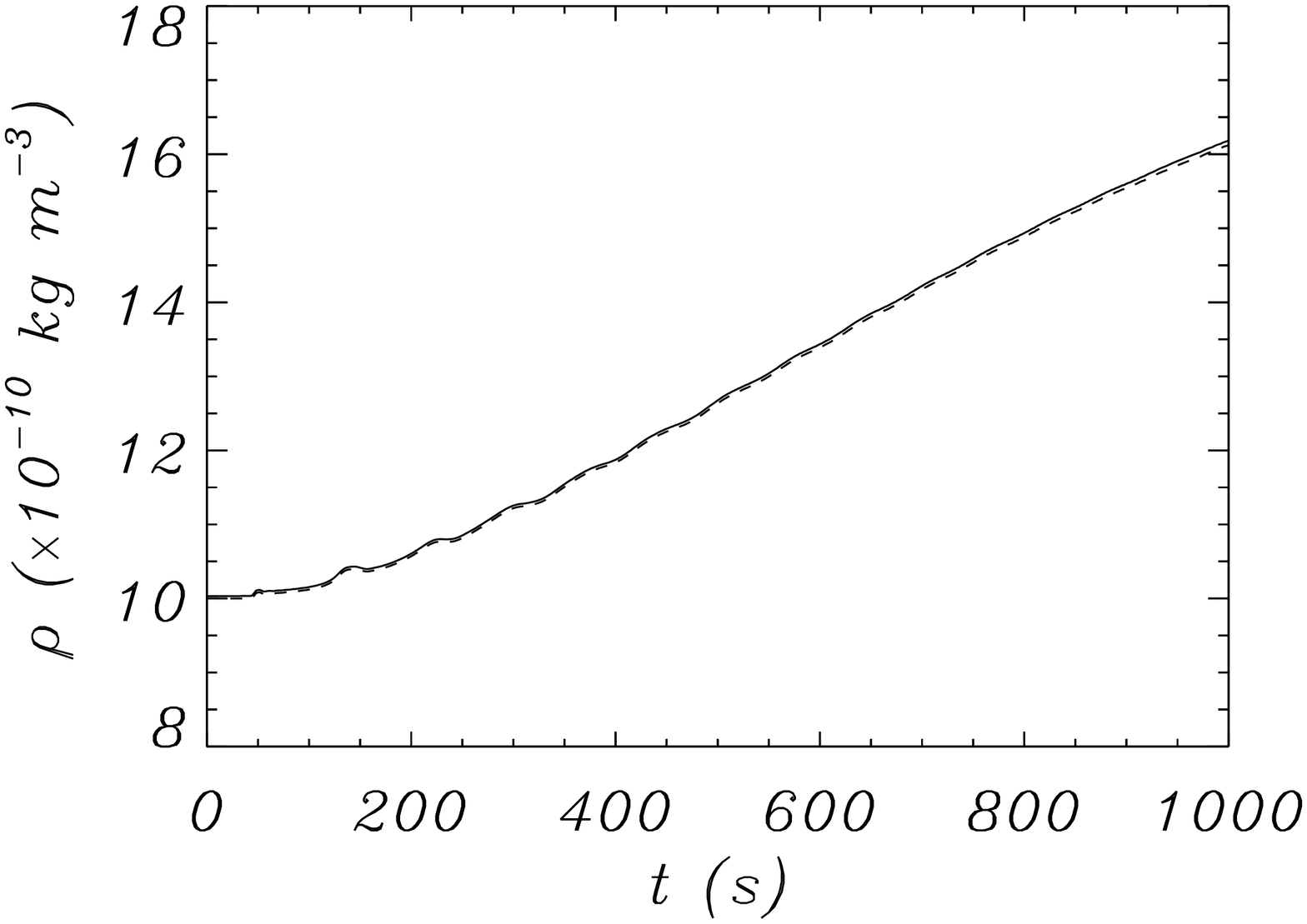} \\
  }
  \centerline{
    \scriptsize{(c)}
    \includegraphics[width=0.33\textwidth,angle=-90]{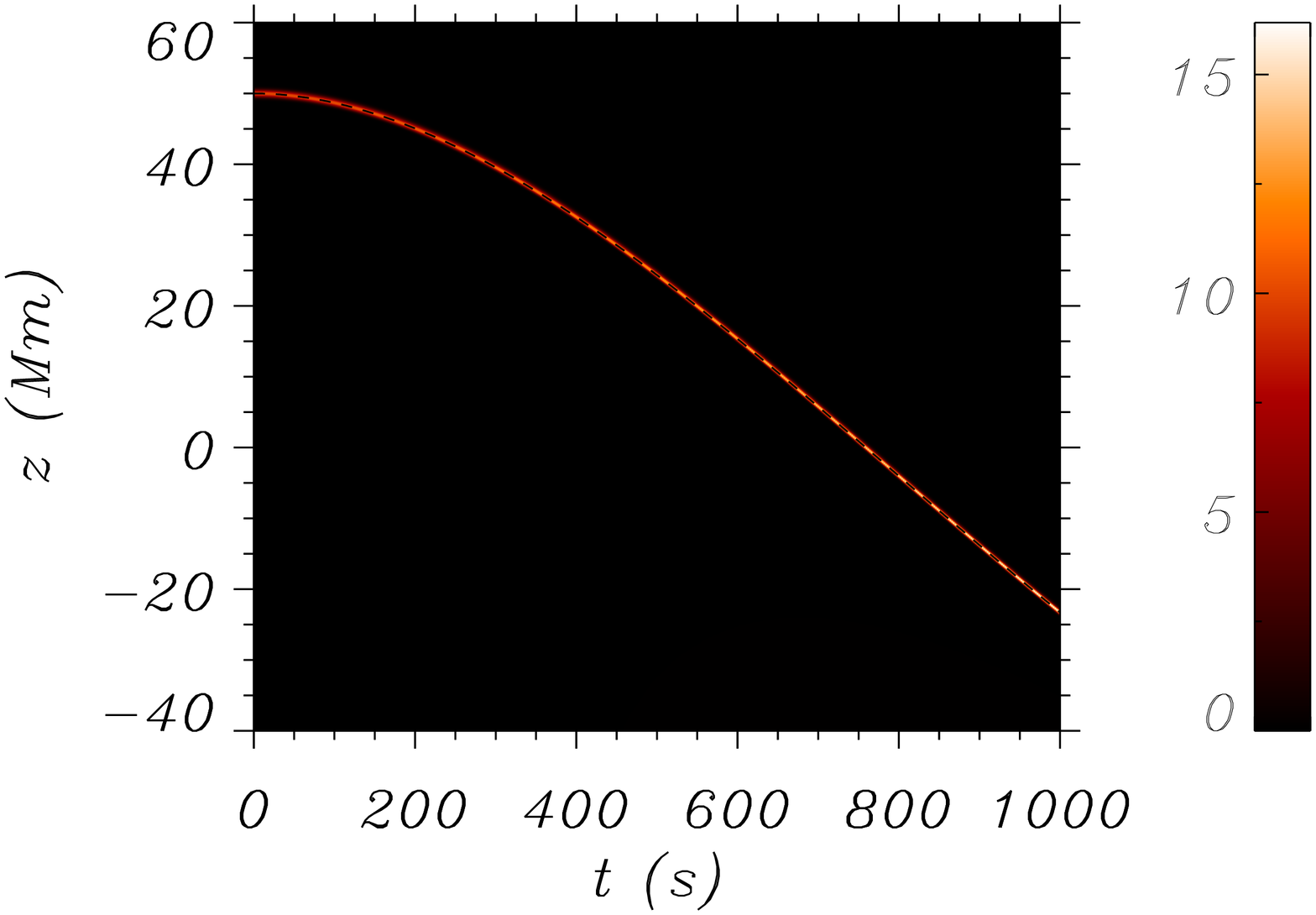}
    \scriptsize{(d)}
    \includegraphics[width=0.33\textwidth,angle=-90]{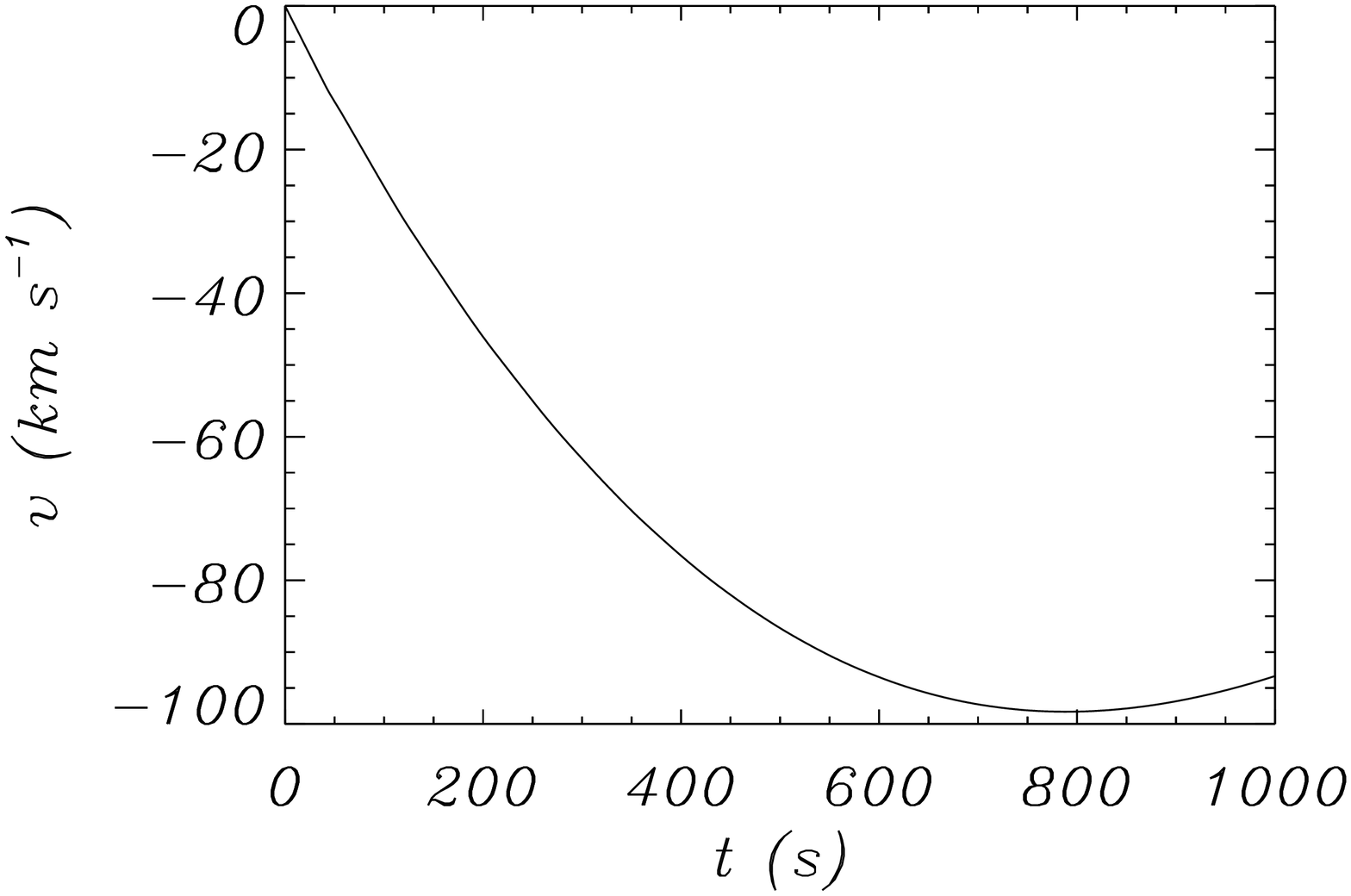}
  }
  \caption{Same as Figure~\ref{fip_density} for a denser mass condensation, namely with maximum value $\rhobiO=10^{-9}$~kg~m$^{-3}$. Note that because of the faster vertical motion the selected times in (a) are not those of Figure~\ref{fip_density}a but $t=0,200,400,600$~s.}
  \label{fip_density10}
\end{figure}

In this section the calculations of \S~\ref{sect_fip_blob_dynamics} are repeated for a blob ten times denser ($\rhobiO=10^{-9}$~kg~m$^{-3}$), while keeping all other parameters unchanged. This heavier blob suffers both a stronger acceleration and a higher density increase. In the previous case (with $\rhobiO=10^{-10}$~kg~m$^{-3}$) the blob descends from $\zO=50$~Mm at $t=0$ to $z\simeq 42$~Mm at $t=600$~s, while in the present case the blob position after the same time is $z\simeq 15$~Mm (Figure~\ref{fip_density10}a). In addition, now the blob density increases by more than 60\%  during the whole simulation (cf. Figure~\ref{fip_density10}b), whereas in the previous (lower density) case the density increase was only about 6\%. The two dynamical stages present in Figure~\ref{fip_density}c (initial acceleration followed by almost constant speed) are not so apparent in the time variation of the density (Figure~\ref{fip_density10}c), but they become clear when we examine its velocity (Figure~\ref{fip_density10}d). In this last plot we see that the acceleration stage lasts much longer in the present case (until $t\simeq 800$~s compared to $t\simeq 250$~s for the less dense blob) and for this reason the maximum descending speed is also much larger ($\vmax\simeq 100$~km s$^{-1}$ compared to $\vmax\simeq 15$~km s$^{-1}$). After this phase, the second stage with a moderate velocity increase is also found.

\begin{figure}[h]
  \centerline{
    \scriptsize{(a)}
    \includegraphics[width=0.33\textwidth,angle=-90]{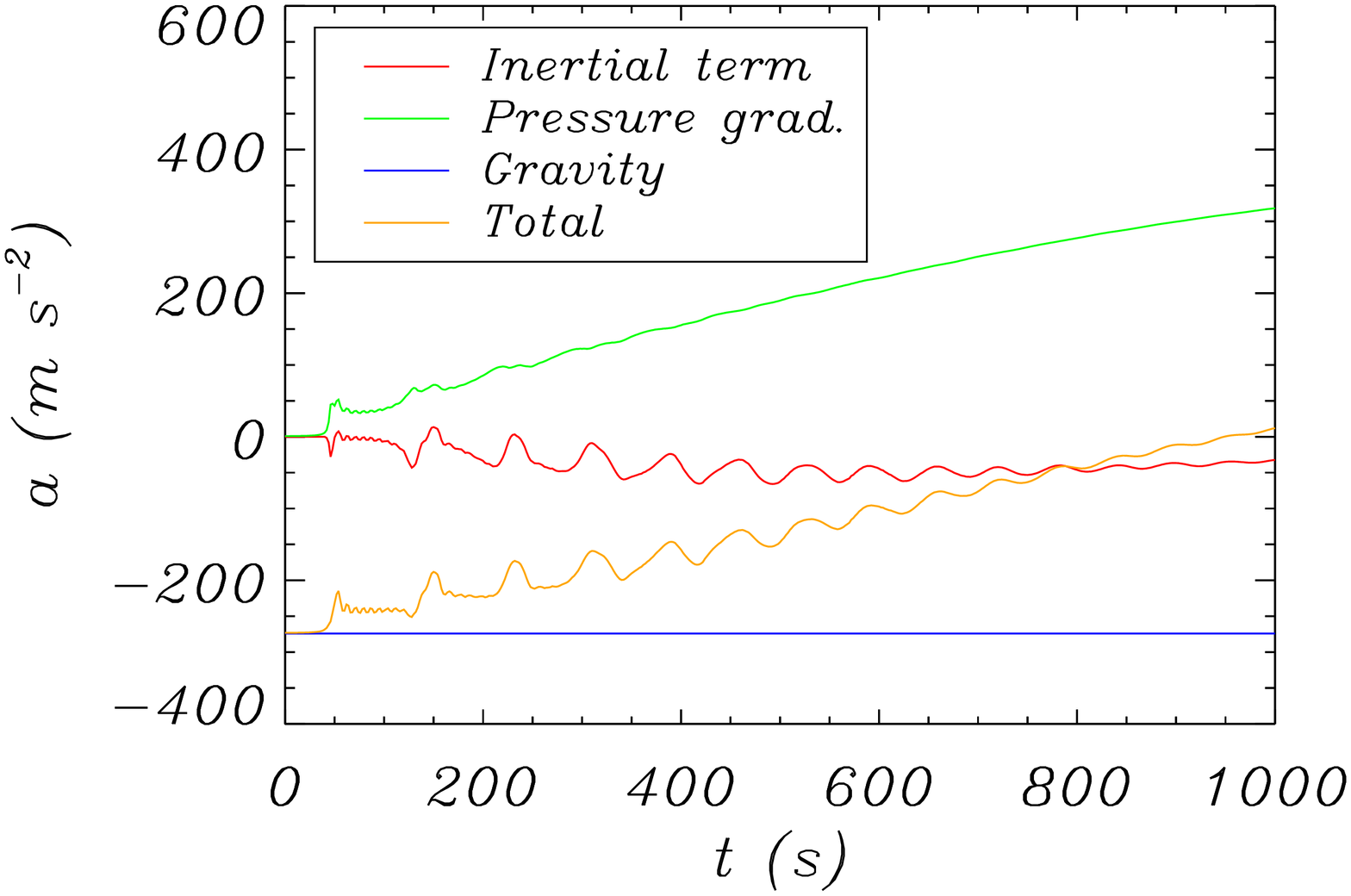}
    \scriptsize{(b)}
    \includegraphics[width=0.33\textwidth,angle=-90]{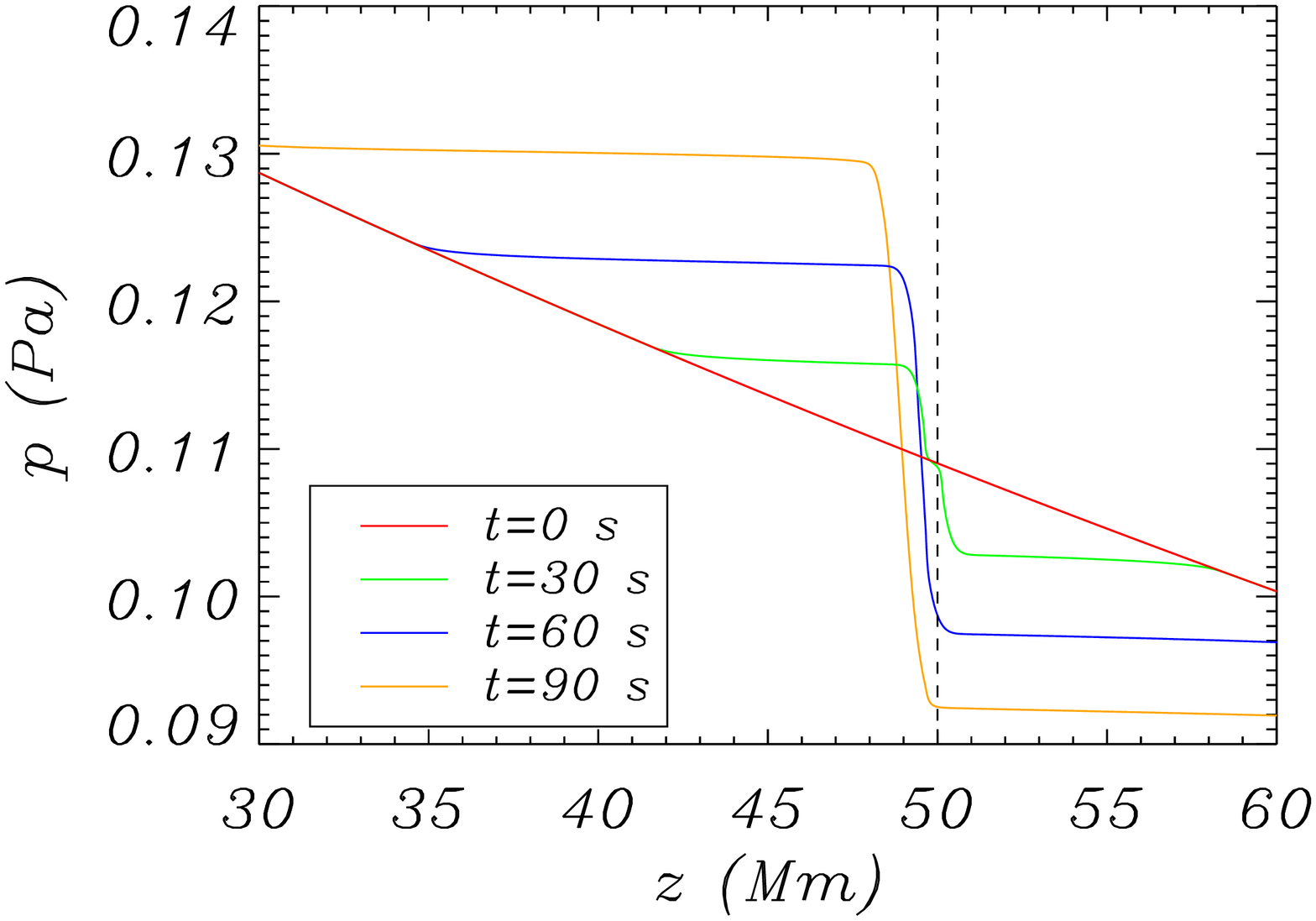}
  }
  \caption{Same as Figure~\ref{fip_dynamics} for a denser mass condensation, namely with $\rhobiO=10^{-9}$~kg~m$^{-3}$.}
  \label{fip_dynamics10}
\end{figure}

Now, the accelerations caused by the forces in the momentum equation (inertial term, pressure gradient, and gravity are represented (Figure~\ref{fip_dynamics10}a). In the present case the blob falls at the acceleration of gravity during the initial $\simeq 40$~s. There are three main differences between Figure~\ref{fip_dynamics10}a and Figure~\ref{fip_dynamics}a. Perhaps the most outstanding one is the presence, during the whole simulation, of wiggles in the inertial term acceleration. The origin of these oscillations is the emission of sound waves by the blob, that is explored in \S~\ref{sect_leaky_waves}. The second difference also concerns the inertial term acceleration, that increases by a factor $\simeq 50$ when $\rhobiO$ is varied from $10^{-10}$~kg~m$^{-3}$ to $10^{-9}$~kg~m$^{-3}$. The reason for this behaviour can be found not only in the increase of the blob velocity (as seen from Figures~\ref{fip_density}d and \ref{fip_density10}d), but also in the increase of its $z$-derivative. And the third difference between Figures~\ref{fip_dynamics}a and \ref{fip_dynamics10}a is that the rate of increase of the pressure gradient is much smaller now and so the total acceleration is negative during a longer time (cf. the two green lines). There are two reasons for this slower build up of the acceleration produced by the pressure gradient. First, to obtain the acceleration caused by $-\nabla\pie$ this gradient is divided by $\rhoi$, which is at least ten times larger for $\rhobiO=10^{-9}$~kg~m$^{-3}$ than for $\rhobiO=10^{-10}$~kg~m$^{-3}$. Second, it is more difficult to generate a steep enough pressure variation at the blob position if it is denser. To explain this point, let us take, for example, the curve in Figure~\ref{fip_dynamics10}b that corresponds to $t=60$~s. The corona under the blob has been disturbed down to a height $z\simeq 35$~Mm by the sound wave excited at the initial blob position ($\zO=50$~Mm), while the pressure distribution below this point is the initial one. Then, the pressure between the blob position and the lowest perturbed height cannot exceed that of this height, which is about 0.124~Pa for $t=60$~s. A similar argument gives the minimum pressure above the blob, $\pie\simeq 0.097$~Pa. The pressure jump between the top and bottom of the blob is limited by the range of heights that has been perturbed at any time, which in turn gives an upper limit to the pressure gradient. The conclusion is that denser blobs require the generation of larger pressure gradients to counteract their weight and so this process (to which we refer as the acceleration phase) lasts longer.

\subsection{Blob temperature}
\label{sect_fip_blob_temp}

Having omitted non-ideal effects in our equations, the blob temperature variation in time is determined purely by the ideal gas law once the plasma pressure and density have been computed. Then, the spatial variation of the temperature in the blob obtained here would probably be quite different had heat conduction (that tends to smear out temperature variations) and radiation (that tends to cool the plasma) been taken into account.

\begin{figure}[h]
  \centerline{
    \scriptsize{(a)}
    \includegraphics[width=0.33\textwidth,angle=-90]{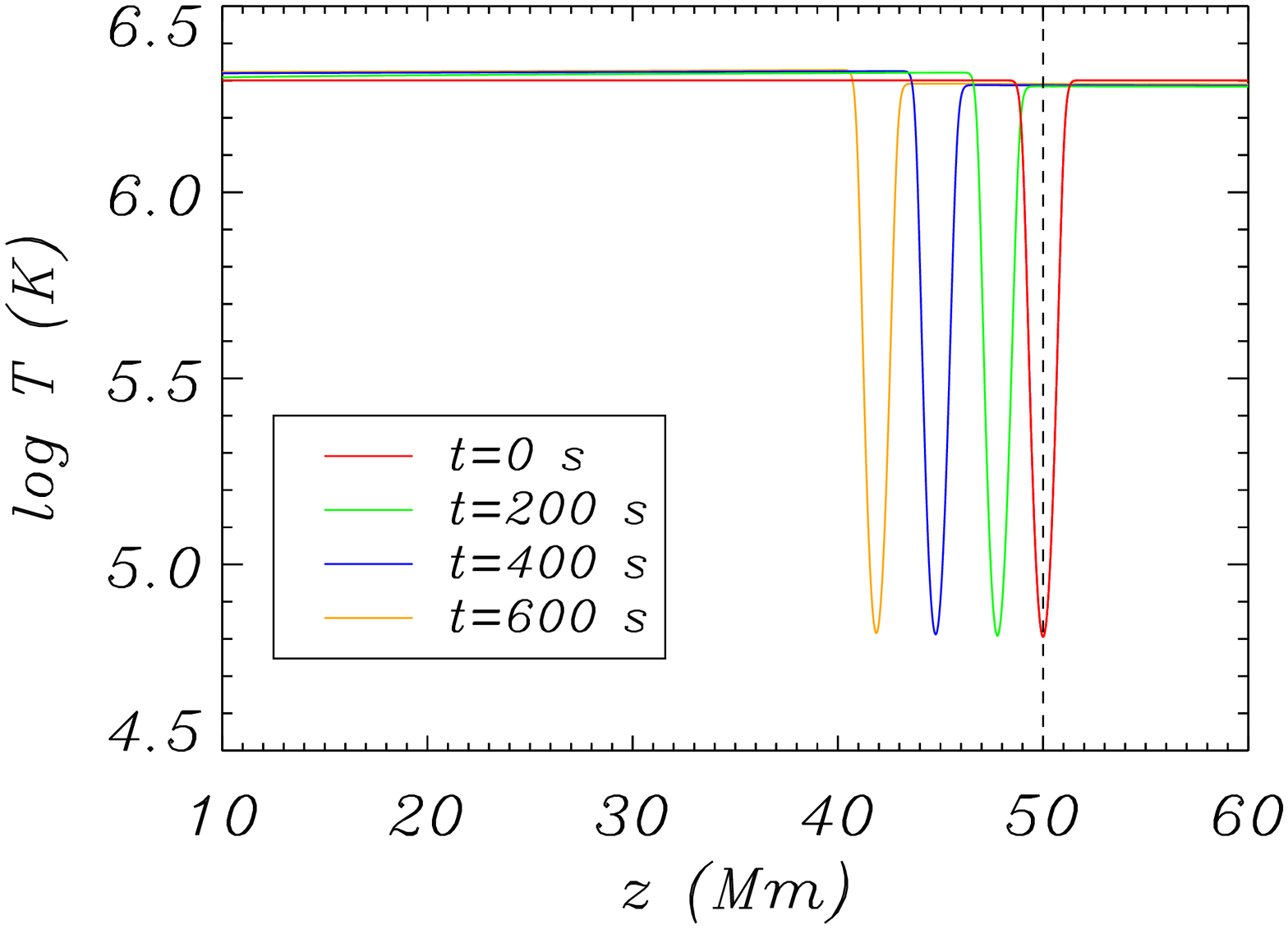}
    \scriptsize{(b)}
    \includegraphics[width=0.33\textwidth,angle=-90]{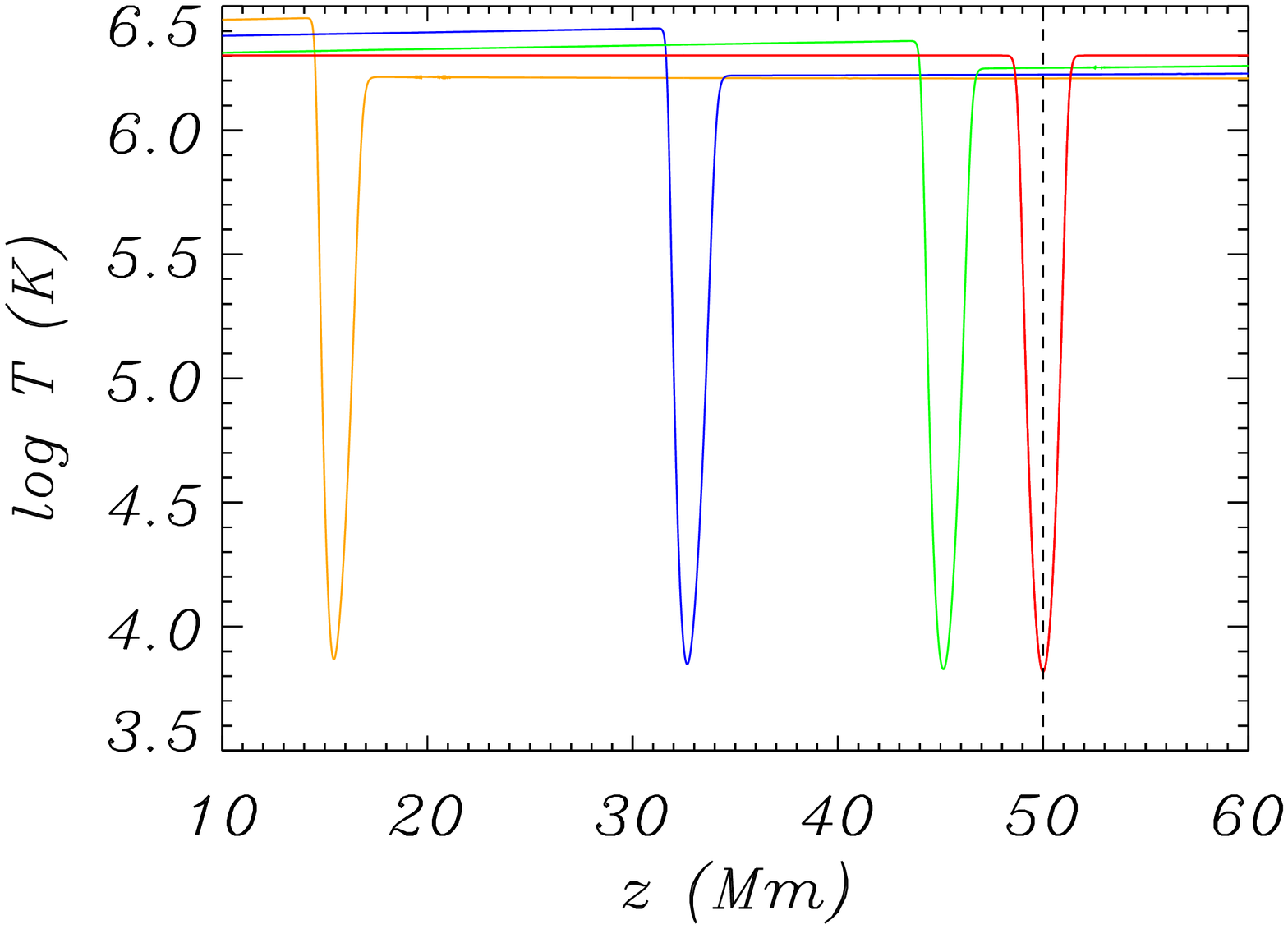}
  }
  \caption{Vertical distribution of the plasma temperature for the times shown in panel (a). The mass condensation has an initial maximum density (a) $\rhobiO=10^{-10}$~kg~m$^{-3}$ and (b) $\rhobiO=10^{-9}$~kg~m$^{-3}$. The vertical line denotes the initial blob position.}
  \label{fip_temperature}
\end{figure}

The temperature of the two simulations of \S~\ref{sect_fip_blob_dynamics} and \ref{sect_fip_blob_dynamics10} are presented in Figure~\ref{fip_temperature} for several times. In the case of the less dense blob (Figure~\ref{fip_temperature}a) the coronal temperature does not depart much from its initial value, $\TO=2\times10^6$~K. The largest differences from the initial temperature are found below the falling condensation and arise because of the pressure restructuring caused by the initial sound wave emission (Figure~\ref{fip_dynamics}a). Moreover, the blob temperature displays a Gaussian shape and shows a minimum value around $\simeq 60.000$~K that remains essentially constant in time. This temperature is higher than expected for coronal rain and descending prominence knots, with observed emission in cool spectral lines. Regarding the simulation with the higher initial blob density (Figure~\ref{fip_temperature}b), now the coronal temperature presents a slight increase below (decrease above) the blob position, caused by a similar variation in the ambient pressure. The blob temperature has a shape similar to that of Figure~\ref{fip_temperature}a and, despite the strong increase of the blob density (see Figure~\ref{fip_density10}a), its minimum temperature only increases slightly during the simulation, namely from $\simeq 7000$~K to $\simeq 8000$~K.

\subsection{Blob shape}
\label{sect_fip_blob_shape}

\begin{figure}[h]
  \centerline{
    \scriptsize{(a)}
    \includegraphics[width=0.33\textwidth,angle=-90]{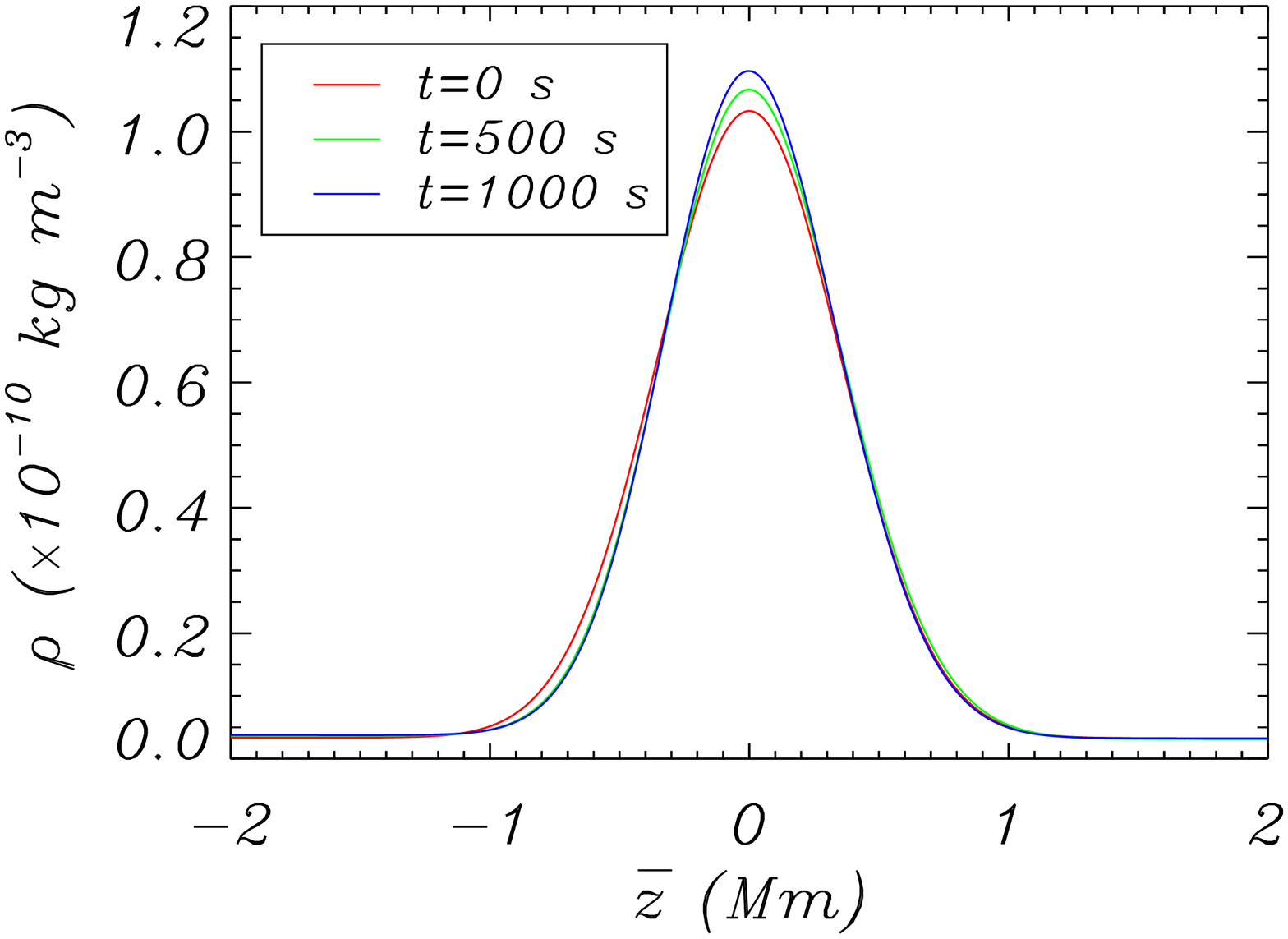} \\
    \scriptsize{(b)}
    \includegraphics[width=0.33\textwidth,angle=-90]{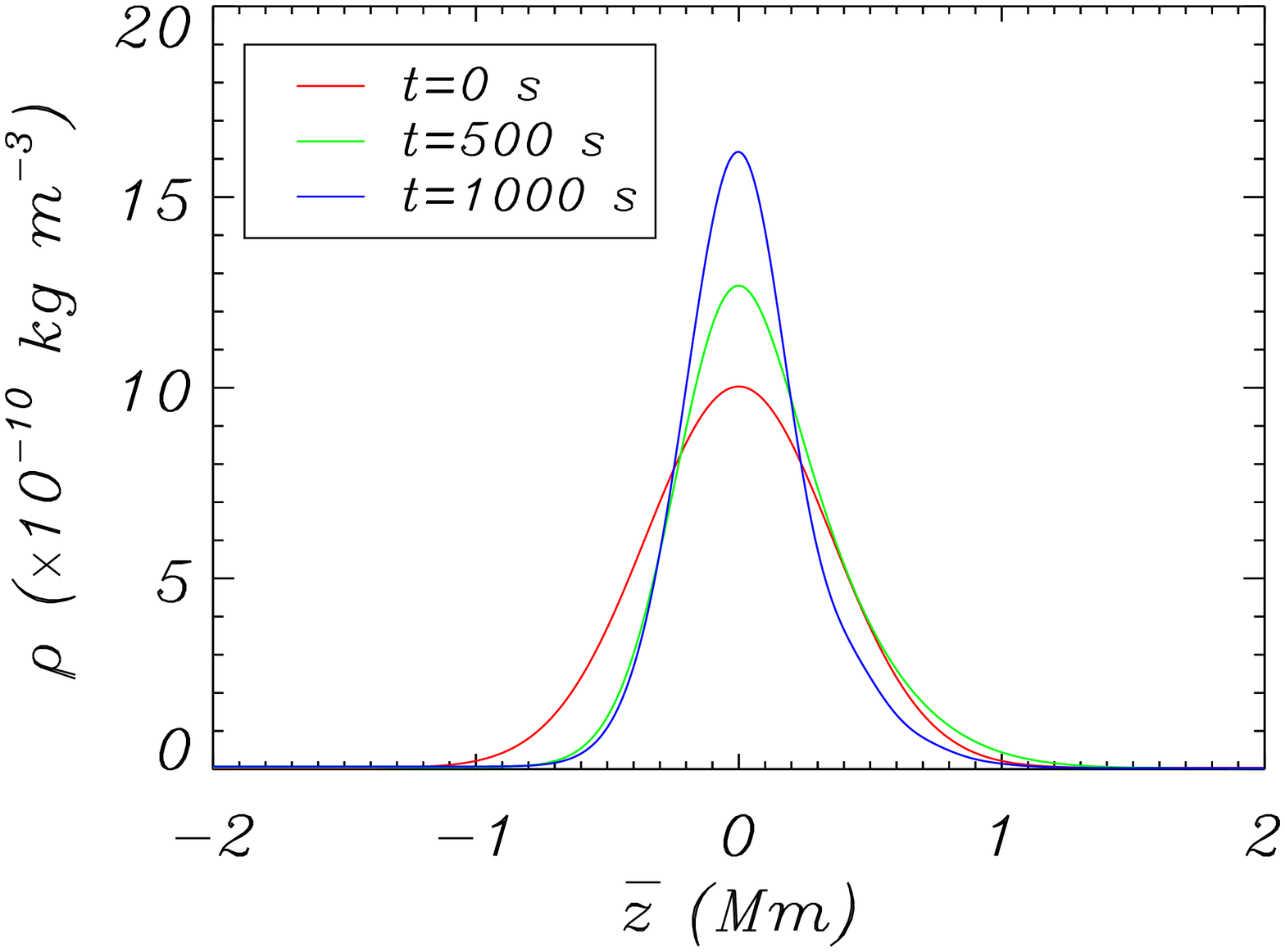} \\
  }
  \caption{Vertical distribution of the blob density for several times. The zero-height level is chosen so that, for each time, the maximum density positions coincide. (a) $\rhobiO=10^{-10}$~kg~m$^{-3}$, (b) $\rhobiO=10^{-9}$~kg~m$^{-3}$. All other parameter values are those of \S~\ref{sect_fip_blob_dynamics} and \ref{sect_fip_blob_dynamics10}. The red curves correspond to the red curves in Figures~\ref{fip_density}a and \ref{fip_density10}a, respectively.}
  \label{fip_blob_shape}
\end{figure}

Here we examine the change in blob shape during the numerical simulations using the results of \S~\ref{sect_fip_blob_dynamics} and \ref{sect_fip_blob_dynamics10}. In Figure~\ref{fip_blob_shape}a we plot together the blob density at $t=0,500,1000$~s with a common height origin. To do this plot, for each time ($t$) a new height variable ($\bar z$) is defined by displacing $z$ so that the height of maximum density is $\bar z=0$. The increase of the maximum density noticed in Figures~\ref{fip_density}a and b is rather evident here and it is accompanied by a slight distortion of the Gaussian shape: after 1000~s the blob density has become a bit steeper in its bottom side ($\bar z<0$) and a bit smoother in its upper side ($\bar z>0$). Since numerical dissipation is negligible in our computations, this seems a real effect. Now, we calculate a proxy for the total blob mass by integrating the density of Figure~\ref{fip_blob_shape}a with respect to $\bar z$ in the range $-2$~Mm~$\leq\bar z\leq 2$~Mm. Given that the environment density increases as the blob descends, the background $\rhoi$ is subtracted before obtaining this integral. The results display a very weak monotonic decrease of the blob mass in time. After 1000~s this mass reduction is about 0.8\% and so this effect does not seem too relevant; we conclude that the blob mass is practically constant during its motion.

A similar plot is presented in Figure~\ref{fip_blob_shape}b for $\rhobiO=10^{-9}$~kg~m$^{-3}$. A comparison of Figures~\ref{fip_blob_shape}a and~b reveals that a denser blob undergoes a larger maximum density increase (as discussed in \S~\ref{sect_fip_blob_dynamics10}) and also a stronger shape distortion, the density at the bottom part of the blob displaying a steep profile at $t=1000$~s. One may wonder whether shocks can develop at the bottom of falling blobs with high enough density. We have found no indication of this phenomenon in observations of rapidly falling cold, dense blobs. In addition, the total blob mass also remains essentially constant in time for the present initial blob density, with a reduction of only 0.5\% with respect to the initial value after 1000~s.

\subsection{Sound wave emission by the blob}
\label{sect_leaky_waves}

\begin{figure}[h]
  \centerline{
    \scriptsize{(a)}
    \includegraphics[width=0.33\textwidth,angle=-90]{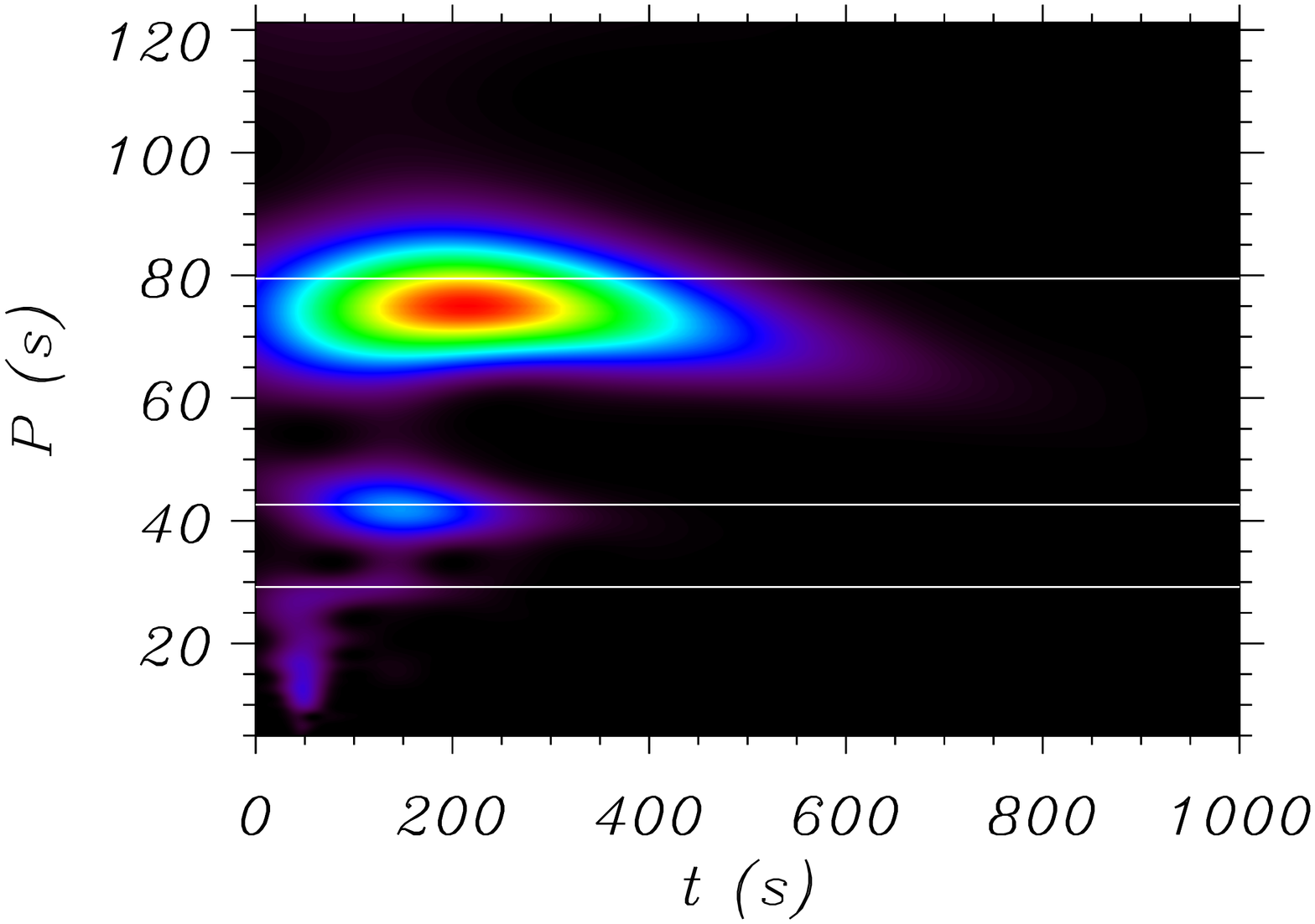}
    \scriptsize{(b)}
    \includegraphics[width=0.33\textwidth,angle=-90]{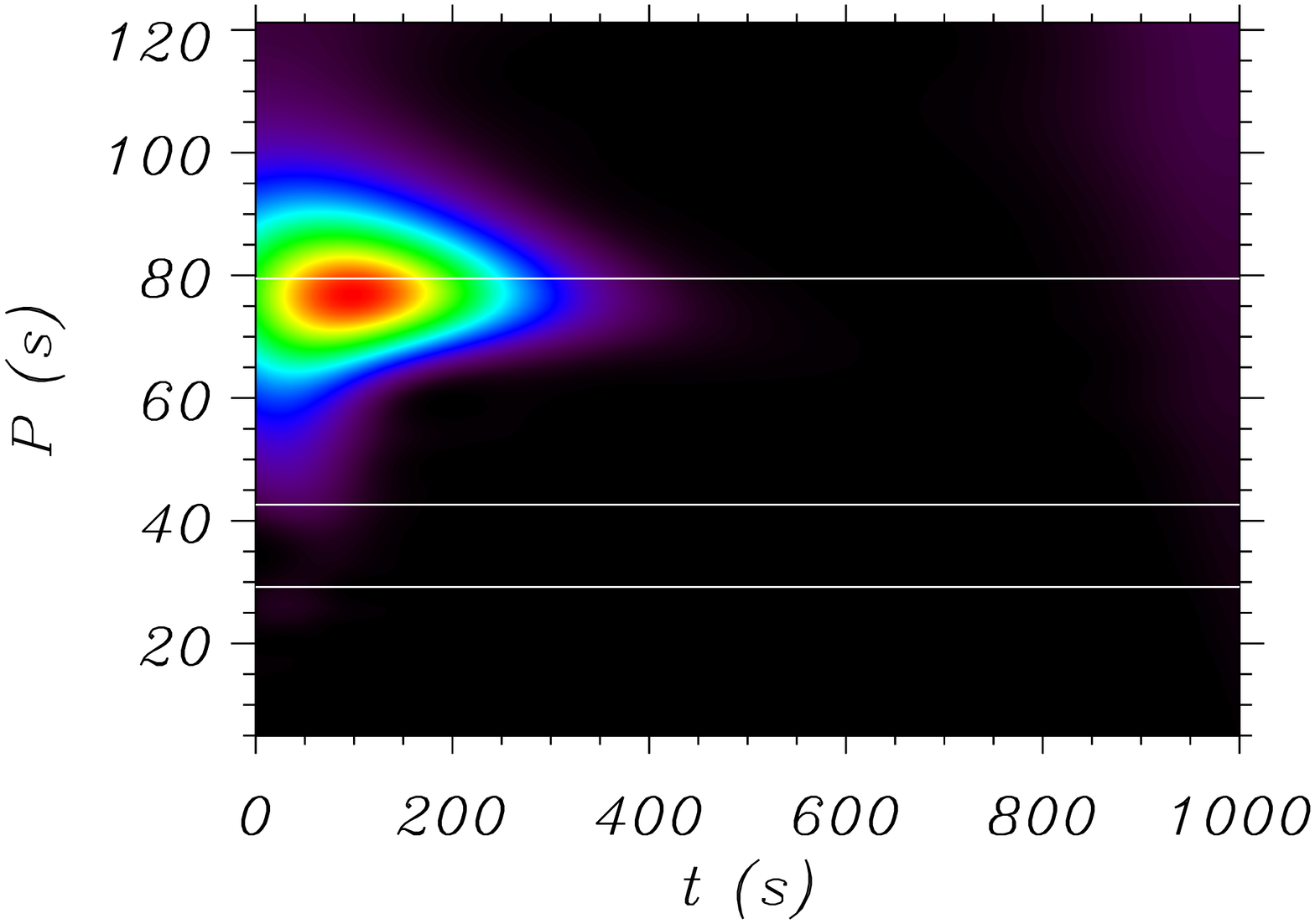} \\
  }
  \centerline{
    \scriptsize{(c)}
    \includegraphics[width=0.33\textwidth,angle=-90]{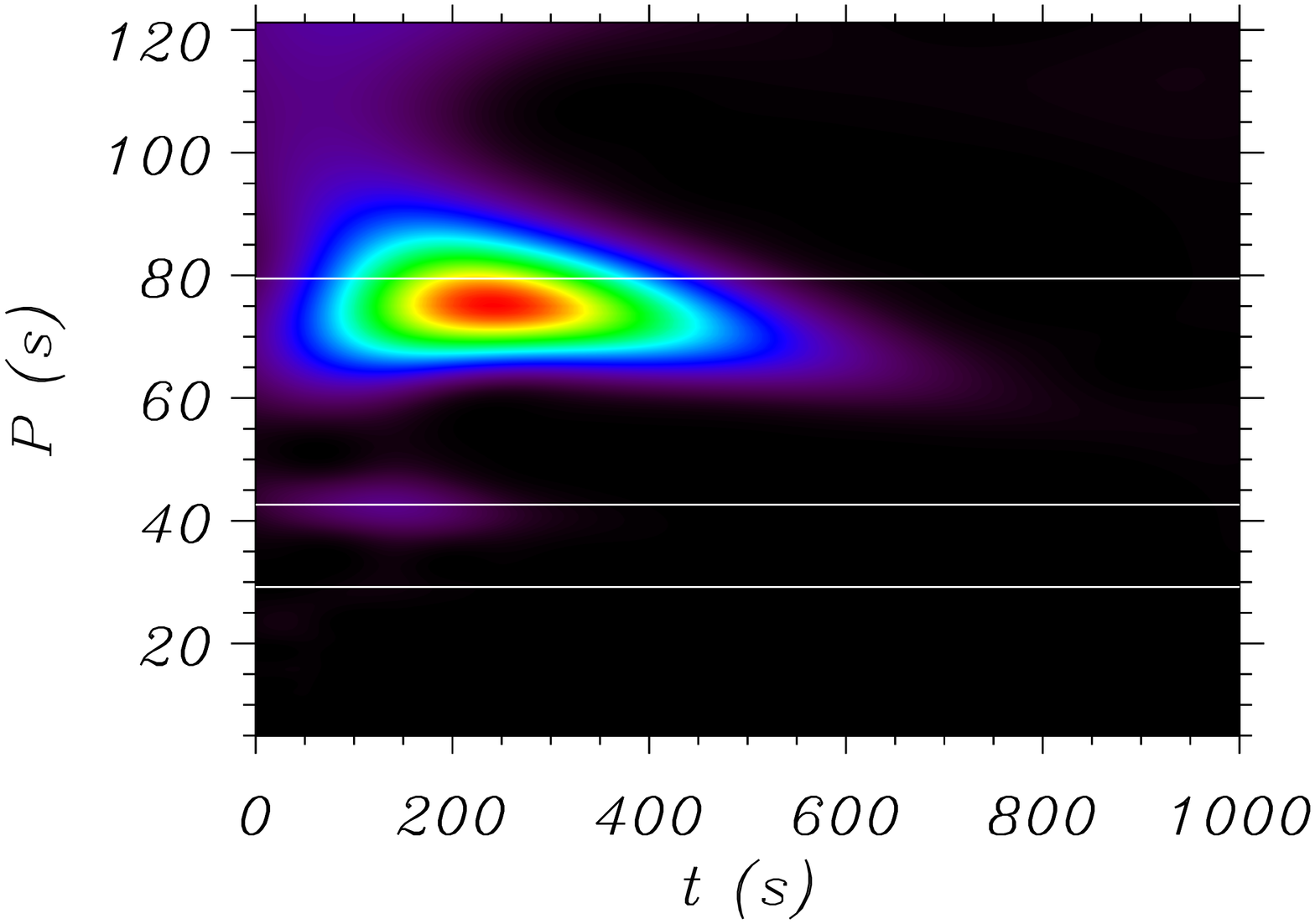} \\
  }
  \caption{Wavelet diagrams of some variables at the blob maximum position for the numerical simulation with initial blob density $\rhobiO=10^{-9}$~kg~m$^{-3}$. (a) Derivative of the vertical velocity with respect to height, (b) vertical velocity, and (c) density. The horizontal and vertical axes give time and period, respectively; wavelet power is in arbitrary units. The horizontal lines correspond to the period of the first three leaky sound waves with even transverse velocity with respect to the blob center (see text). The Morlet wavelet with the parameter $\omega_0=10$ has been used \citep[see][for more details]{torrence98}.}
  \label{fip_wavelet}
\end{figure}

We next pay attention to the outstanding oscillations of the inertial term acceleration (that is equal to $-\vi\partial\vi/\partial z$) seen in Figure~\ref{fip_dynamics10}a. Note that this acceleration (like all others in this figure) is taken at the position of the blob maximum density and so these oscillations must come from periodic variations in $\vi$ and/or $\partial\vi/\partial z$ at the maximum blob position. To investigate the presence of periodicities in these signals we have made use of the wavelet analysis, that has also been applied to the blob maximum density (Figure~\ref{fip_density10}b). Before applying the wavelet analysis, the trends of $\vi$ and $\rhoi$ have been removed by fitting and subtracting a third degree polynomial. We start with $\partial\vi/\partial z$, that has the strongest oscillatory power. Its wavelet diagram (Figure~\ref{fip_wavelet}a) displays three periodicities with periods $P\simeq 75$~s, $P\simeq 42$~s, and $P\simeq 28$~s together with power at the full range of periods between 10 and 30~s. We concentrate in the three features with well-defined periods. They correspond to the oscillations visible in Figure~\ref{fip_dynamics10}a and have three properties: their period remains essentially constant in time, they have small amplitude, and the oscillatory power decays in time, with shorter periods displaying a quicker attenuation. The vertical velocity (Figure~\ref{fip_wavelet}b) only shows the first of these periodicities, while the blob density contains the first and second ones (Figure~\ref{fip_wavelet}c). Hence, the blob is subject to some phenomenon that produces at least two or three periodic, damped variations of $\partial\vi/\partial z$, $\vi$, and $\rhoi$.

To understand these damped oscillations we notice that the blob has smaller sound speed than its environment. This sound speed depression is similar to the Alfv\'en speed depression of a coronal magnetic slab, a structure that supports both trapped and leaky fast waves \citep[e.g.][]{terradas05}. We can then establish an analogy between the two cases and so we hypothesise that the falling blob can support small amplitude, leaky sound waves. This analogy is developed in Appendix~\ref{sect_appendix}. The first three leaky sound waves whose transverse velocity is even about the blob center have periods $P=79.5$~s, 42.6~s, and 29.2~s, that agree quite well with those in the numerical simulations (see horizontal lines in Figure~\ref{fip_wavelet}). Moreover, the corresponding exponential damping times of these modes are $\taud=291$~s, 182~s, and 135~s, so that our finding that shorter periodicities damp faster is also in agreement with these theoretical leaky mode results.

The calculations carried out in Appendix~\ref{sect_appendix} give us more information than just the wave frequency. They indicate that solutions with even velocity about the blob center have non-zero velocity together with vanishing density perturbation and derivative of the velocity at the blob center (i.e., they are vertical blob oscillations that do not change its shape). This is in contradiction with our finding that the three quantities oscillate during the course of the blob descent. There are two reasons for this discrepancy: first, there is a continuous flow of material from the lower part of the blob into it and out from the blob above it. This issue is presented in \S~\ref{sect_fip_plasma_elements} and could be incorporated in Appendix~\ref{sect_appendix} by assuming an equilibrium flow, i.e., a non-zero velocity in Equation~(\ref{wave_equil}). In these circumstances, we do no longer have even or odd solutions about the blob center and all three variables ($\vi$, $\partial\vi/\partial z$, $\rhoi$) are non-zero there. This symmetry is also broken by the change of the blob shape, that is more important for larger blob densities. And finally, the simple equilibrium used in Appendix~\ref{sect_appendix} is just a coarse approximation to the dynamical situation of a falling blob.

The results discussed in this section correspond to the numerical simulation with the densest blob. In the case of the numerical simulation with a smaller initial blob density the inertial term does not seem to have the same type of oscillations (cf. Figure~\ref{fip_dynamics}a); we nevertheless have repeated the wavelet analysis with the same three variables ($\vi$, $\partial\vi/\partial z$, and $\rhoi$ at the blob maximum). The results are analogous for the two simulations, although those of the simulation with the largest initial density ($\rhobiO=10^{-9}$~kg~m$^{-3}$) display clearer oscillatory features. So we conclude that the larger the blob density, the stronger the sound wave emission during its fall. In addition, the initial blob density is also crucial both in the period and damping time of the oscillations, that are both smaller for the less dense blob: $P=25.8$~s, 13.9~s, 9.54~s and $\taud=30.0$~s, 18.7~s, 13.6~s. The ratio of damping time to period determines the number of periods required to achieve a given attenuation. This quantity is much smaller for the simulation with initial blob density $\rhobiO=10^{-10}$~kg~m$^{-3}$ than for that with $\rhobiO=10^{-9}$~kg~m$^{-3}$. In other words, the less dense blob has oscillations with smaller amplitude that attenuate much faster, so the detection of this sound wave emission during the blob fall is a signature of a high density. The reason for the smaller periods and shorter attenuation times in the less dense blob must be found in the larger value of the sound speed in the blob: this reduces the transit time inside the blob (and hence the period) and makes the sound speed depression shallower (and so worsens the blob capability for trapping sound waves).

\subsection{Motion of plasma elements}
\label{sect_fip_plasma_elements}

\begin{figure}[ht!]
  \centerline{
    \scriptsize{(a)}
    \includegraphics[width=0.33\textwidth,angle=-90]{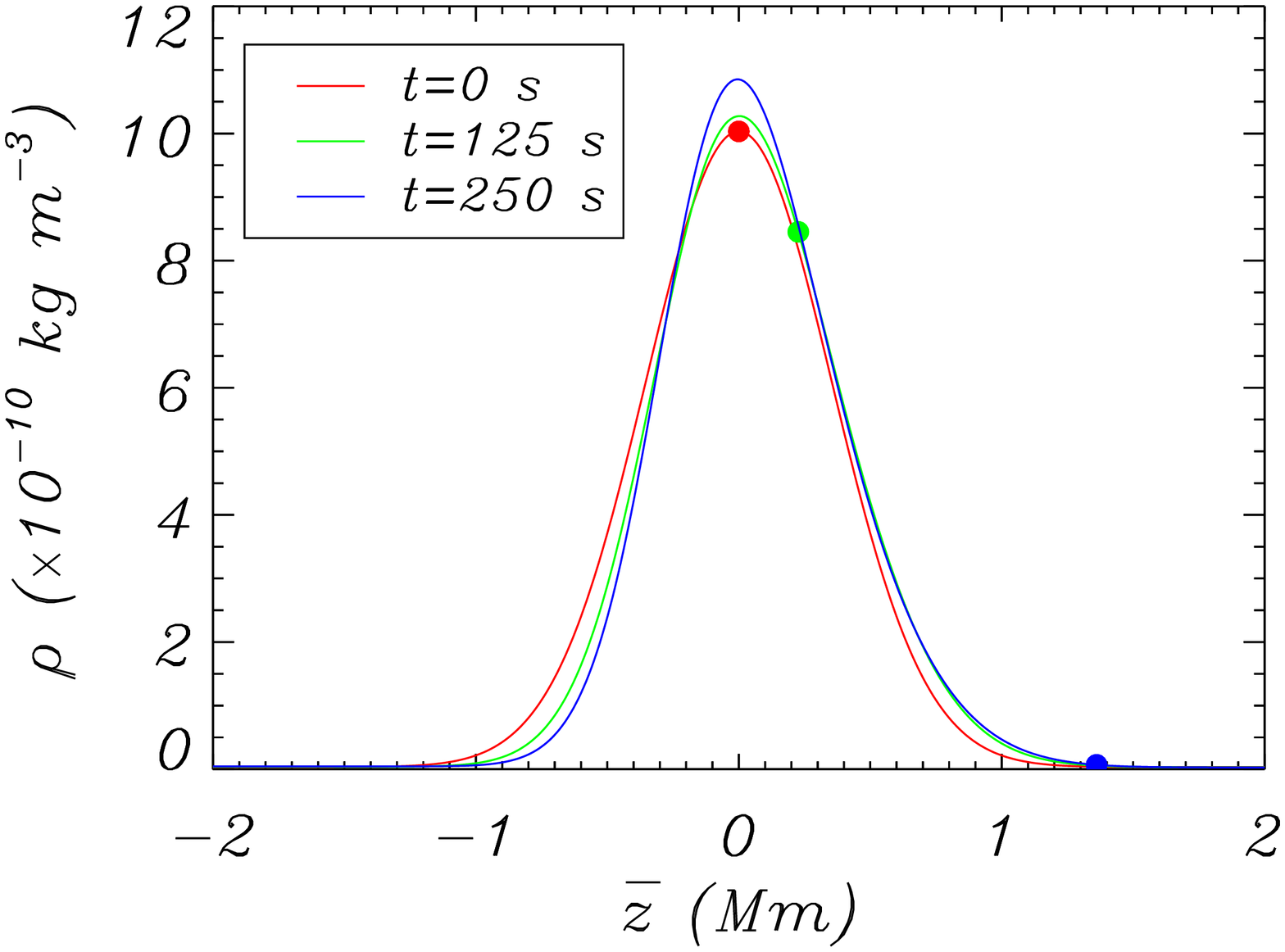} \\
    \scriptsize{(b)}
    \includegraphics[width=0.33\textwidth,angle=-90]{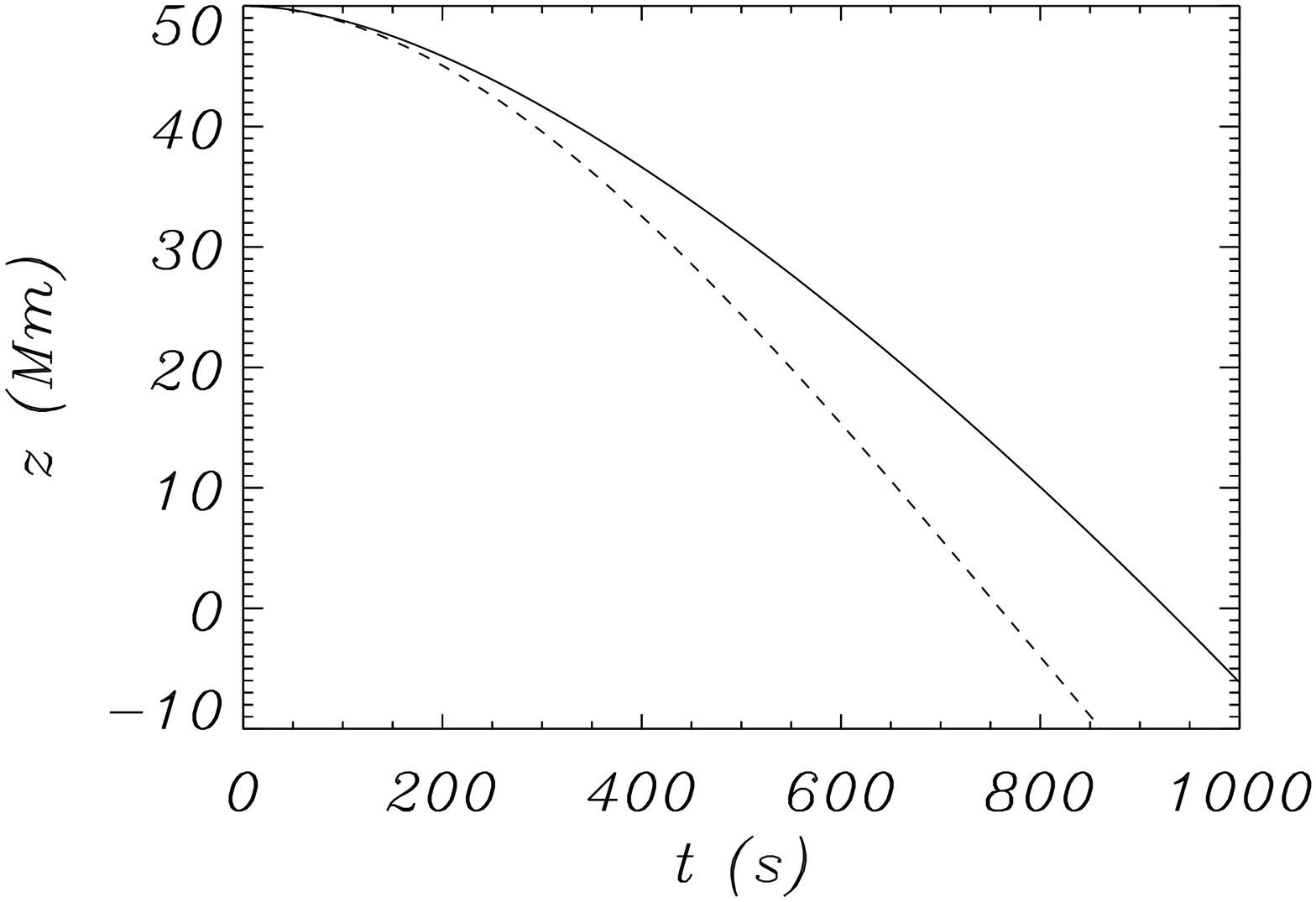} \\
  }
  \centerline{
    \scriptsize{(c)}
    \includegraphics[width=0.33\textwidth,angle=-90]{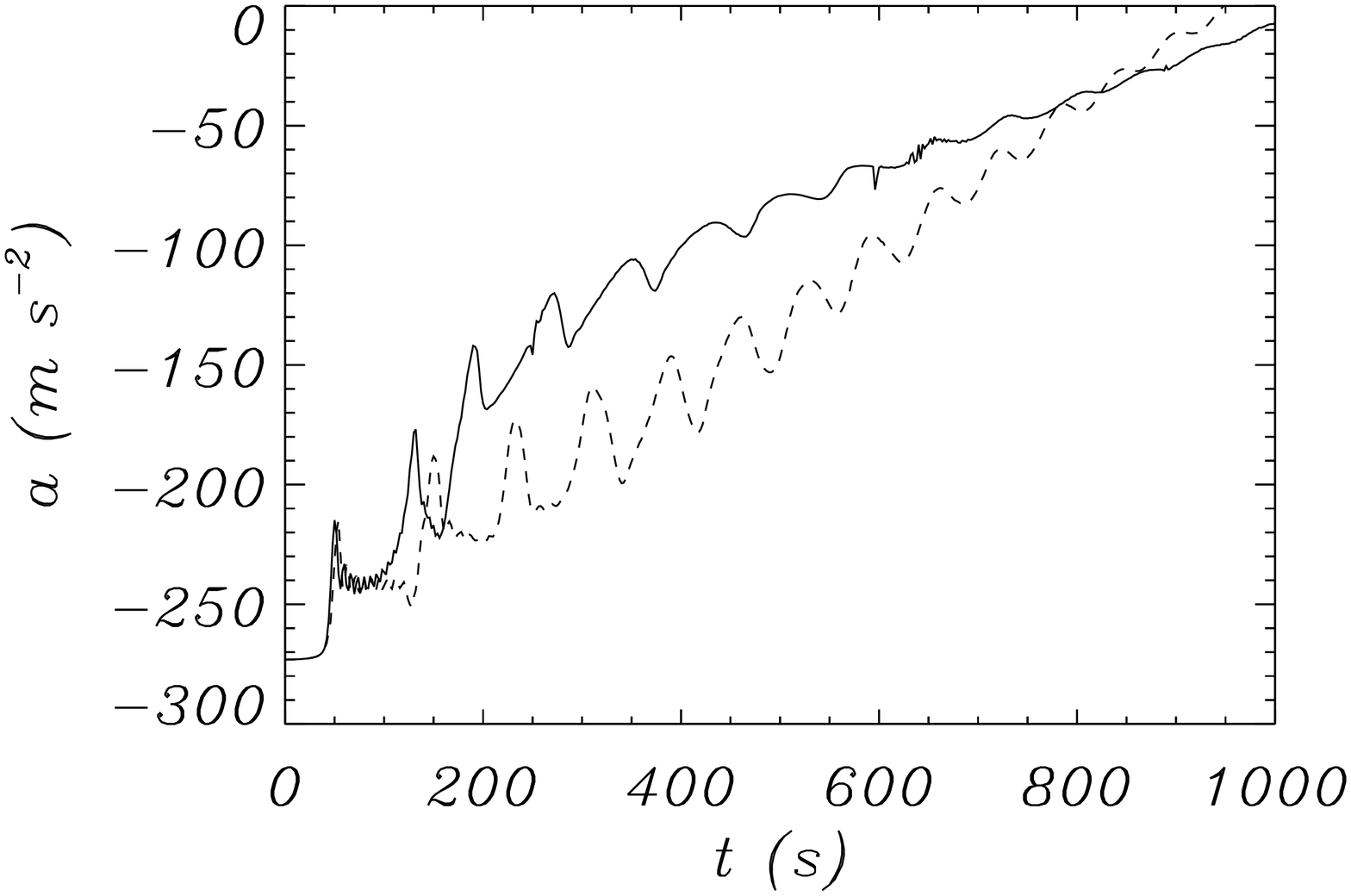} \\
  }
  \caption{(a) Same as Figure~\ref{fip_blob_shape}b for $t=0,125,250$~s, that is, vertical distribution of the blob density for several times with the height origin placed at the maximum density position. The initial blob maximum density is $\rhobiO=10^{-9}$~kg~m$^{-3}$. For each time, the position of the fluid element with the maximum initial density is given by a filled circle. (b) Height and (c) acceleration of the fluid element with the maximum initial density (solid) and of the maximum density position (dashed). The dashed curves in panels (b) and (c) correspond to the dashed curve of Figure~\ref{fip_density10}c and the orange curve of Figure~\ref{fip_dynamics10}a, respectively.}
  \label{fip_plasma_element}
\end{figure}

The motion of each fluid element can be followed in time thanks to the Lagrangian displacement, $\xii(z,t)$. An element of plasma initially at a height $\zini$ will be at a height $\zini+\xii(\zini,t)$ at time $t$. Here the results for $\rhobiO=10^{-9}$~kg~m$^{-3}$ are used to discuss the behavior of the plasma element initially at the blob center, i.e., with initial height $\zini=50$~Mm. Figure~\ref{fip_plasma_element}a reveals that this element moves upwards relative to the blob maximum and that after 250~s it is even outside the blob. Thus, as the blob falls down, plasma below it is engulfed by the condensation (and therefore highly compressed) and, at the same time, plasma in the blob is left above it and so must expand to meet the local, coronal density. The continuous separation between the plasma element initially at $z=50$~Mm and the blob maximum position is better seen by representing their respective positions (Figure~\ref{fip_plasma_element}b), that slowly diverge in time. The total acceleration of these two plasma elements is shown in Figure~\ref{fip_plasma_element}c. During most of the time, the blob maximum position has a downward acceleration that is larger by a few tens of m~s$^{-2}$, and this causes the departure displayed in Figures~\ref{fip_plasma_element}a and b. All the effects discussed in this paragraph are also found for the smaller blob initial density ($\rhobiO=10^{-10}$~kg~m$^{-3}$), although with a lower magnitude.

\subsection{Parametric study}
\label{sect_fip_parametric}

\begin{figure}[h]
  \centerline{
    \scriptsize{(a)}
    \includegraphics[width=0.33\textwidth,angle=-90]{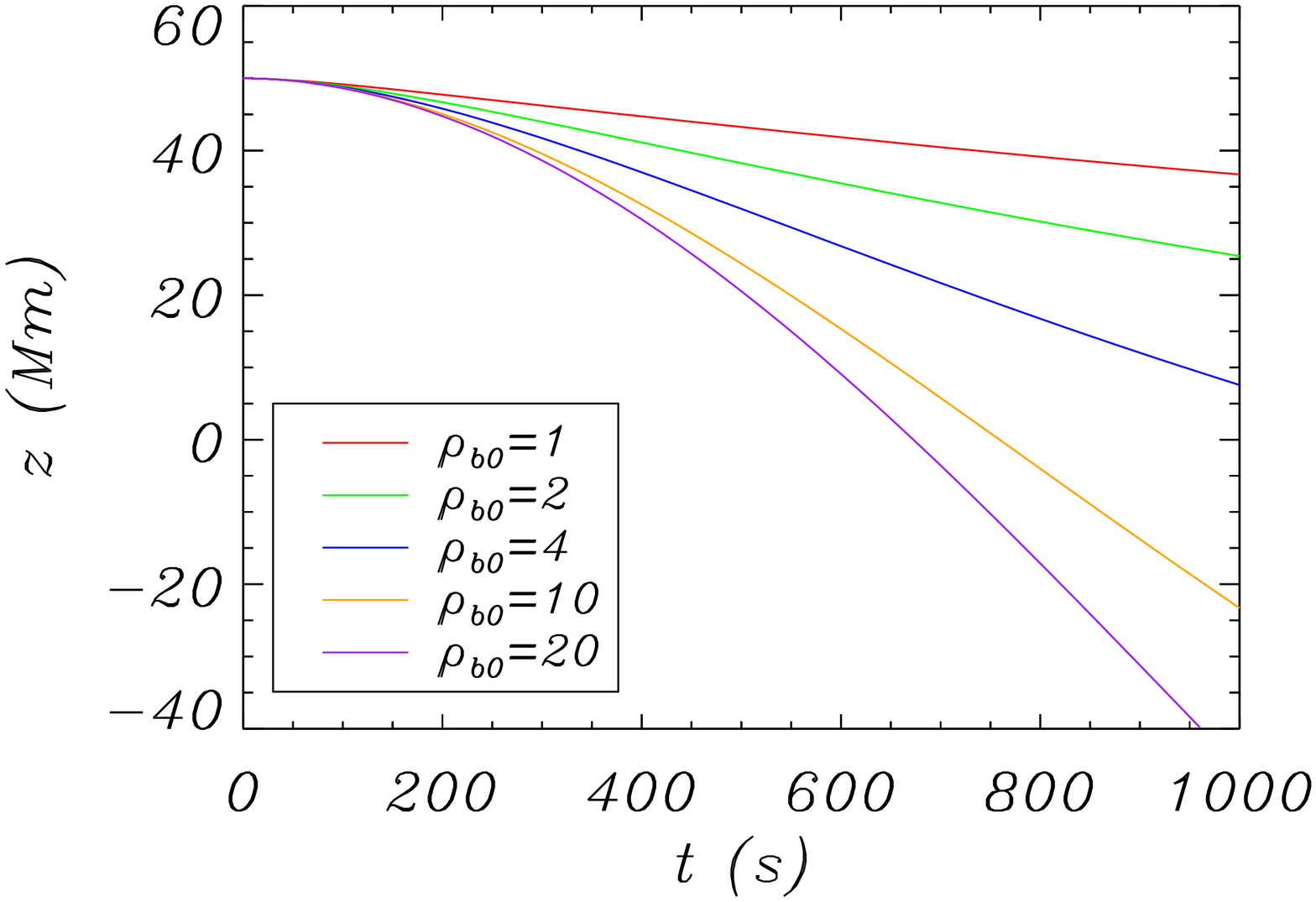}
    \scriptsize{(b)}
    \includegraphics[width=0.33\textwidth,angle=-90]{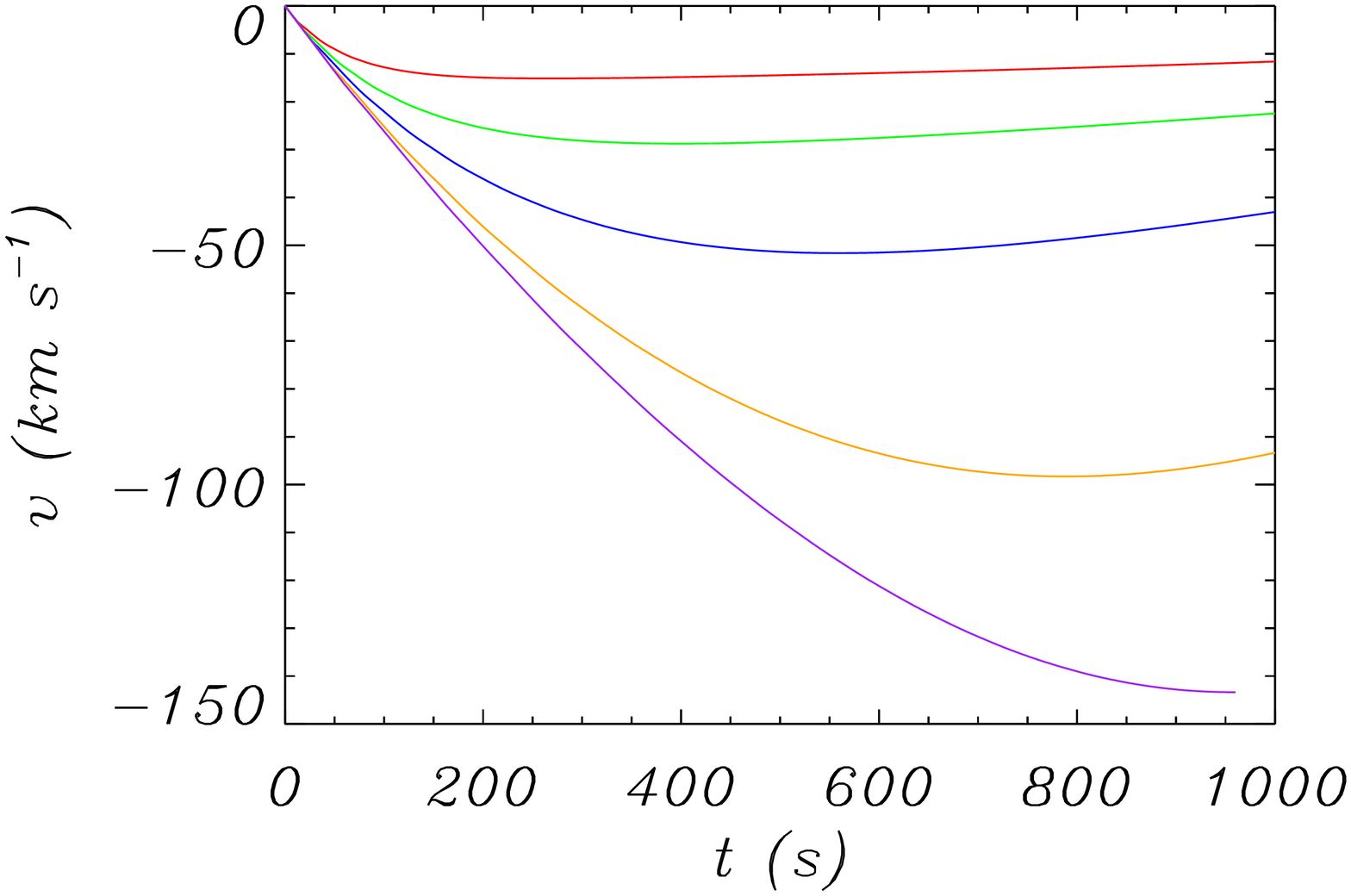}
  }
  \centerline{
    \scriptsize{(c)}
    \includegraphics[width=0.33\textwidth,angle=-90]{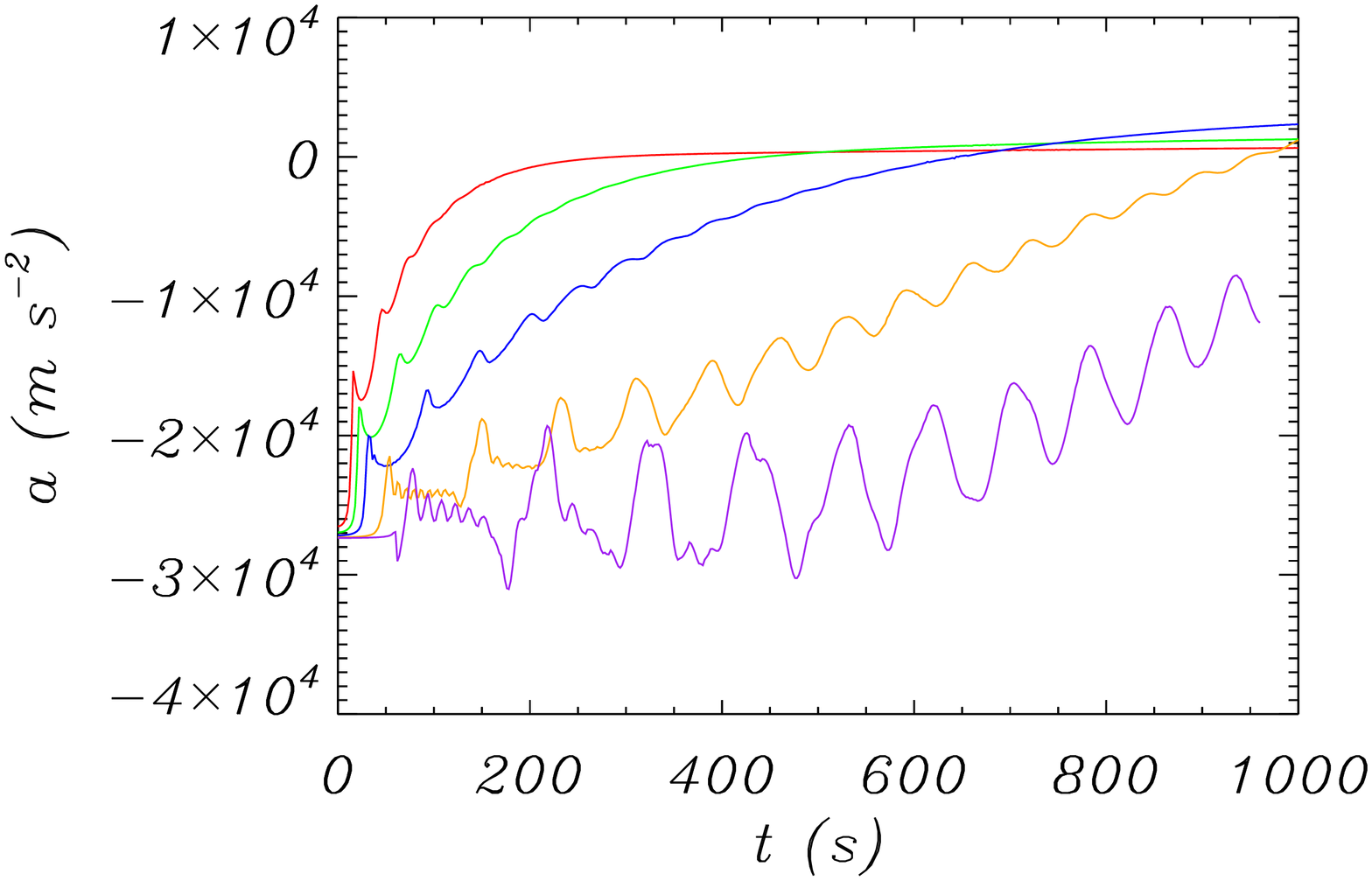}
  }
  \caption{(a) Blob height, (b) blob speed, and (c) blob acceleration as a function of time. The parameter values are those of \S~\ref{sect_fip_blob_dynamics} and \ref{sect_fip_blob_dynamics10} except for the blob maximum density, $\rhobiO$, whose value is given in panel (a) in units of $10^{-10}$~kg~m~$^{-3}$. The red (orange) lines have already been presented in Figures~\ref{fip_density} and \ref{fip_dynamics} (\ref{fip_density10} and \ref{fip_dynamics10}).}
  \label{fip_t0_2d6}
\end{figure}

In the numerical simulation presented above with $\rhobiO=10^{-10}$~kg~m$^{-3}$ the blob speed is considerably smaller than that of many coronal rain blobs and descending prominence knots. On the other hand, the falling speed attained by the blob with initial mximum density $\rhobiO=10^{-9}$~kg~m$^{-3}$ is considerably larger and surpasses that of descending prominence knots and even some coronal rain events. Our purpose now is to determine the influence of the parameter values on the mass condensation dynamics and on its maximum descending speed. We consider as reference values those used in \S~\ref{sect_fip_blob_dynamics} and \ref{sect_fip_blob_dynamics10} and change only one parameter at a time.

First the blob maximum density, $\rhobiO$, is varied in the range $10^{-10}-2\times 10^{-9}$~kg~m$^{-3}$. The blob height versus time is presented in Figure~\ref{fip_t0_2d6}a. In all cases the blob displays the initial acceleration stage followed by a period with a more or less straight trajectory, that is, a more or less constant speed. This figure illustrates well our previous finding that denser blobs fall faster, such as can be confirmed from the blob velocity, see Figure~\ref{fip_t0_2d6}b. We note that consecutive values of $\rhobiO$ differ more or less by a factor of two and that in these two panels the curves for $\rhobiO=10^{-9}$~kg~m$^{-3}$ and $\rhobiO=2\times 10^{-9}$~kg~m$^{-3}$ are closer than other adjacent pairs of curves. Hence, it seems that increasing the initial blob density even more will lead to a more moderate modification of the blob trajectory and maximum descending speed. Both Figures~\ref{fip_t0_2d6}b and c confirm that the initial acceleration phase is present for all blob densities, although it lasts longer for larger densities of the mass condensation. In addition, in all cases except that of the densest condensation, the blob always has an upward acceleration at the end of the simulation. Finally, Figure~\ref{fip_t0_2d6}c also confirms that denser blobs are characterised by larger oscillations in their acceleration, i.e., stronger sound wave emission.

\begin{figure}[ht]
  \centerline{
    \scriptsize{(a)}
    \includegraphics[width=0.33\textwidth,angle=-90]{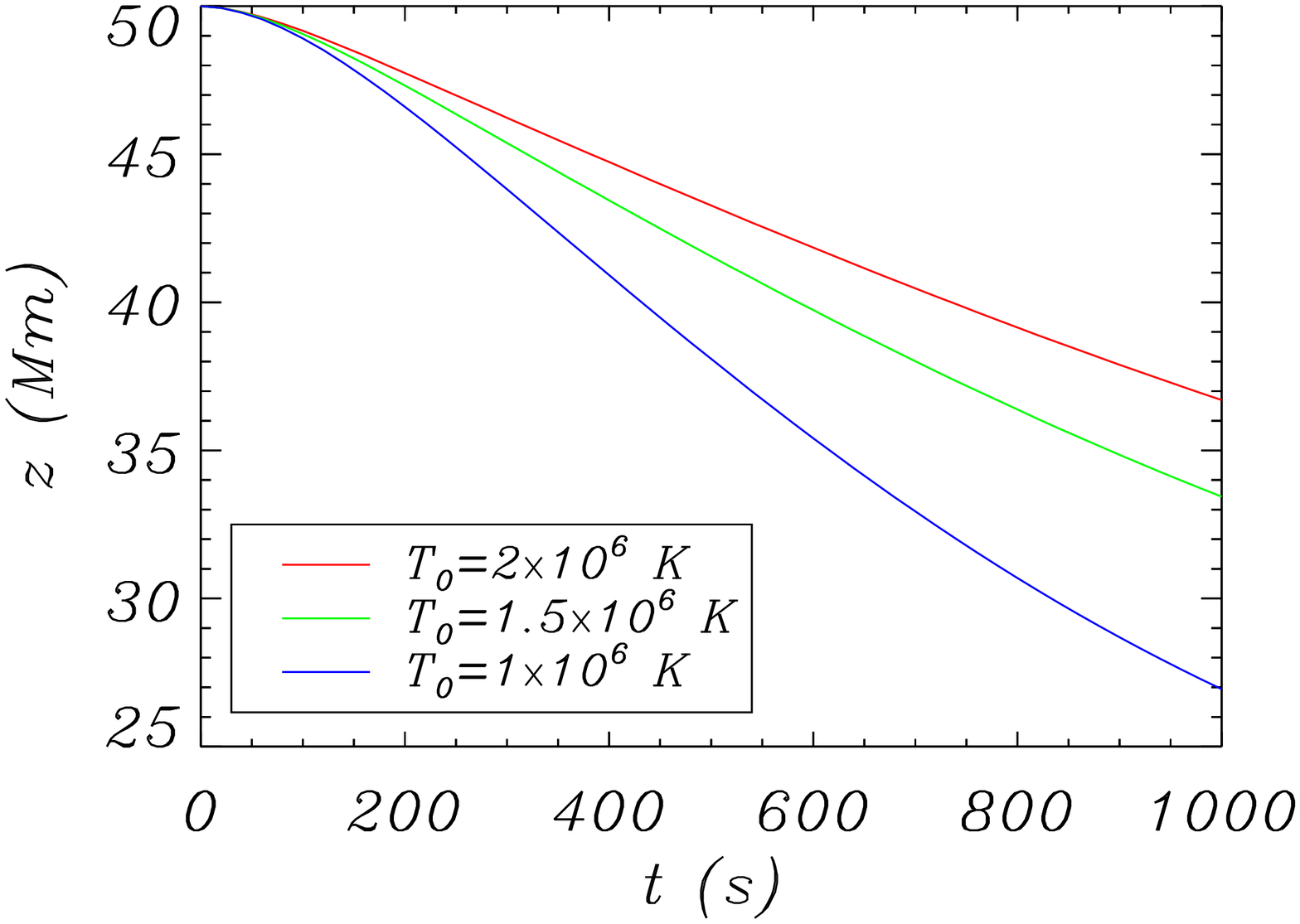}
    \scriptsize{(b)}
    \includegraphics[width=0.33\textwidth,angle=-90]{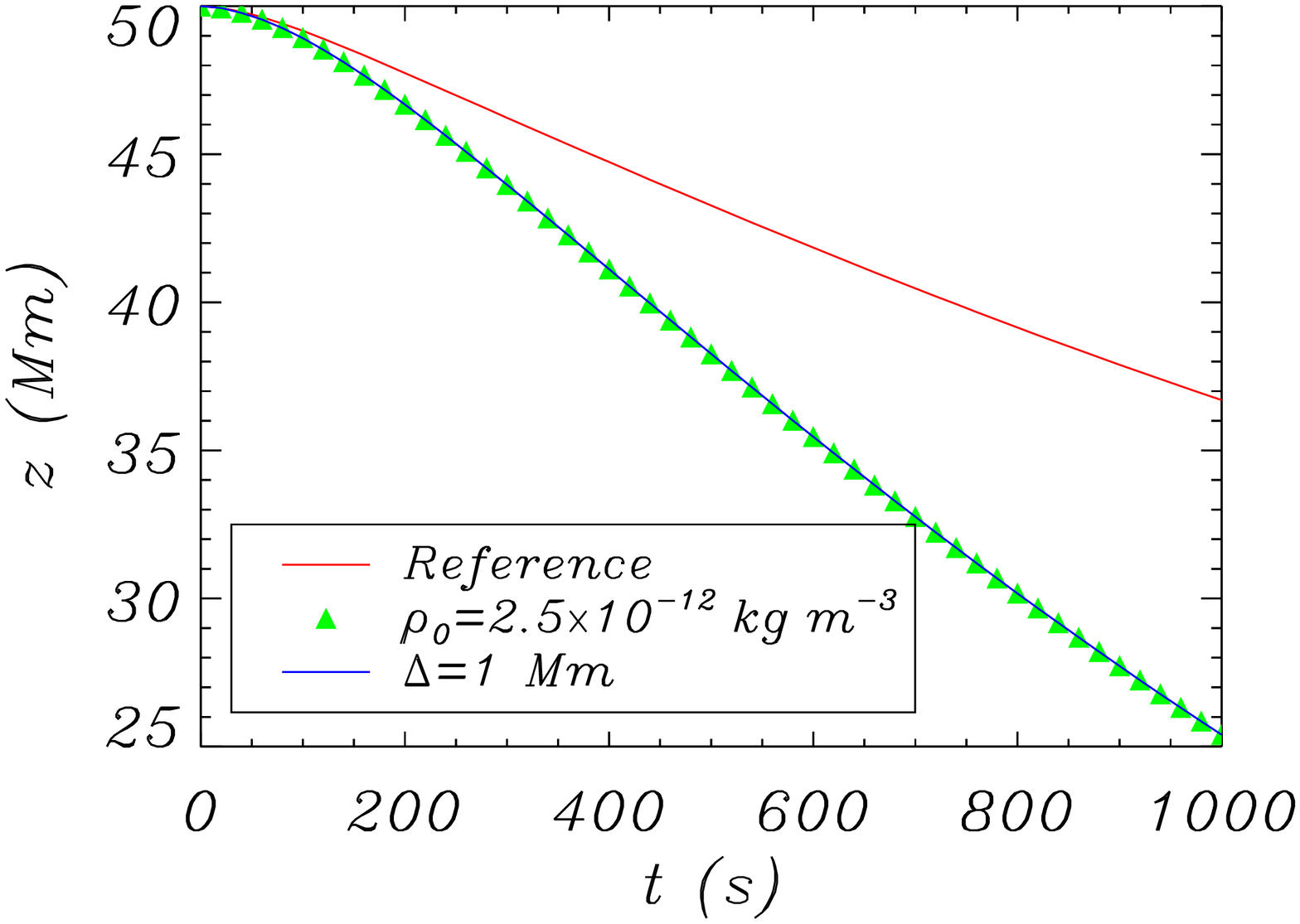}
  }
  \caption{Blob height versus time for (a) various coronal temperatures and (b) a different coronal base density ($\rhoiO$ is divided by two with respect to the reference value) or blob length ($\Delta$ is multiplied by two with respect to the reference value). Only one of these parameters ($\TO$, $\rhoiO$, $\Delta$) has been changed at a time and so all other parameter values are the reference ones of \S~\ref{sect_fip_blob_dynamics}. Then, the red line in the two panels corresponds to the curve of Figure~\ref{fip_density}c and the red curve of Figure~\ref{fip_t0_2d6}a.}
  \label{fip_param}
\end{figure}

Since the height dependence on time gives enough information on the blob kinematics, now we do not show the blob velocity and acceleration. First the influence of the background temperature is studied: along with the reference value $\TO=2\times 10^6$~K, we have also considered $\TO=1\times 10^6$~K and $\TO=1.5\times 10^6$~K (Figure~\ref{fip_param}a). Varying the coronal temperature does not modify the shape of the trajectory, but a reduction in $\TO$ leads to a descent with higher downward acceleration. This effect is explained as follows: smaller values of $\TO$ lead to a decrease of the vertical stratification scale height and, given that the base density has not been altered, this implies that the blob moves in an ambient plasma with reduced density. Thus, the blob to environment density ratio is larger. This is analogous to maintaining the coronal structure and increasing the blob density, such as in Figure~\ref{fip_t0_2d6}, and so both a blob with higher density and a corona with smaller temperature result in a stronger acceleration and a larger descending speed. As a conclusion, the ratio of blob to background density controls the blob descent.

Now, a similar effect is obtained if the background temperature is maintained and the base coronal density (and hence plasma pressure) is decreased: the blob also falls in a rarer environment and achieves greater speeds (see triangles in Figure~\ref{fip_param}b). Moreover, if the blob length is doubled, while the rest of parameters are held fixed, then its mass doubles and once more it falls at a larger speed (Figure~\ref{fip_param}b). In fact, this figure shows that halving the coronal density or doubling the blob size produce the same height versus time variation. These two blob trajectories are also identical to that of the case with double initial blob density (green line in Figure~\ref{fip_t0_2d6}a).

\begin{figure}[ht]
  \centerline{
    \scriptsize{(a)}
    \includegraphics[width=0.33\textwidth,angle=-90]{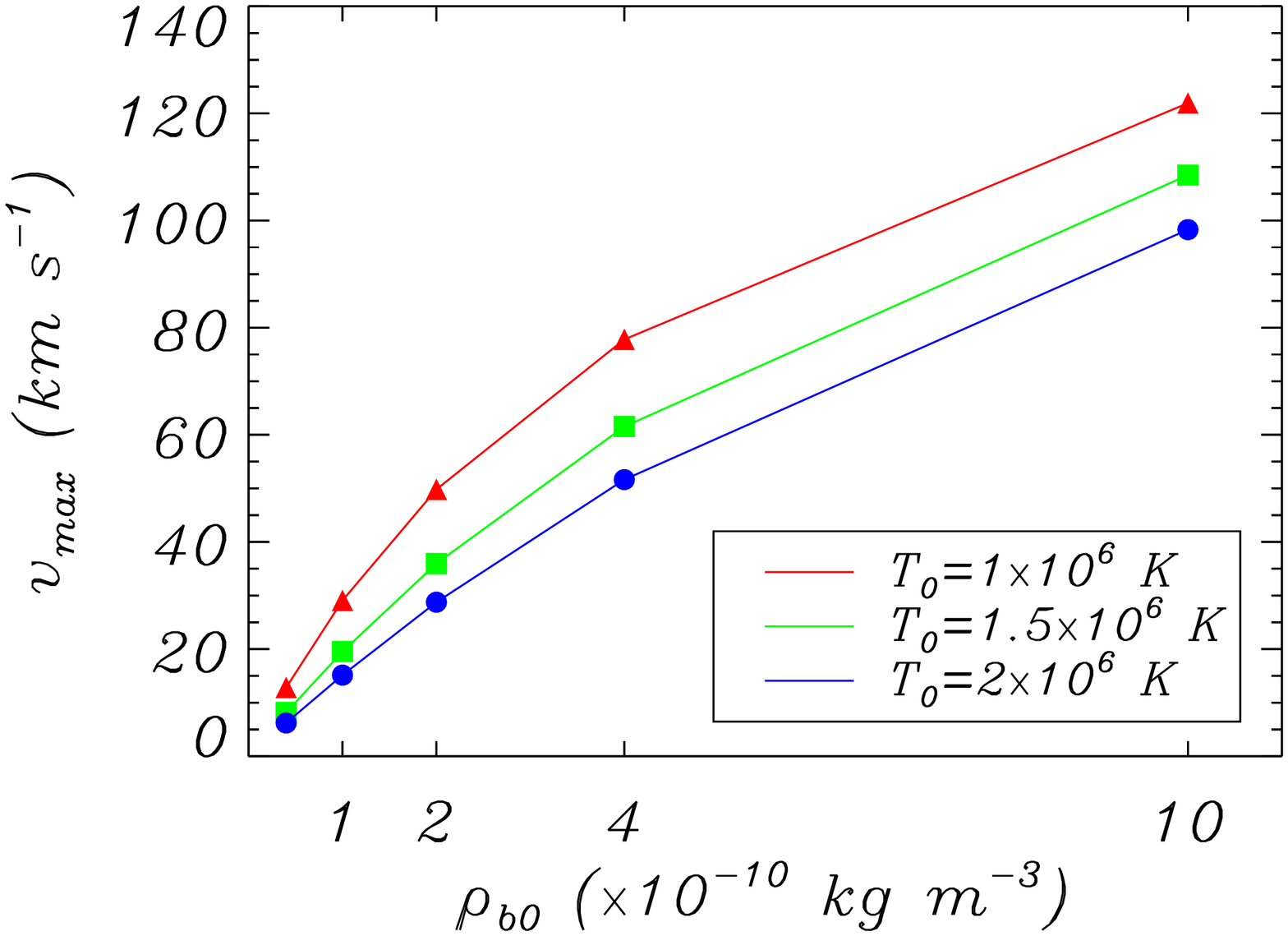}
    \scriptsize{(b)}
    \includegraphics[width=0.33\textwidth,angle=-90]{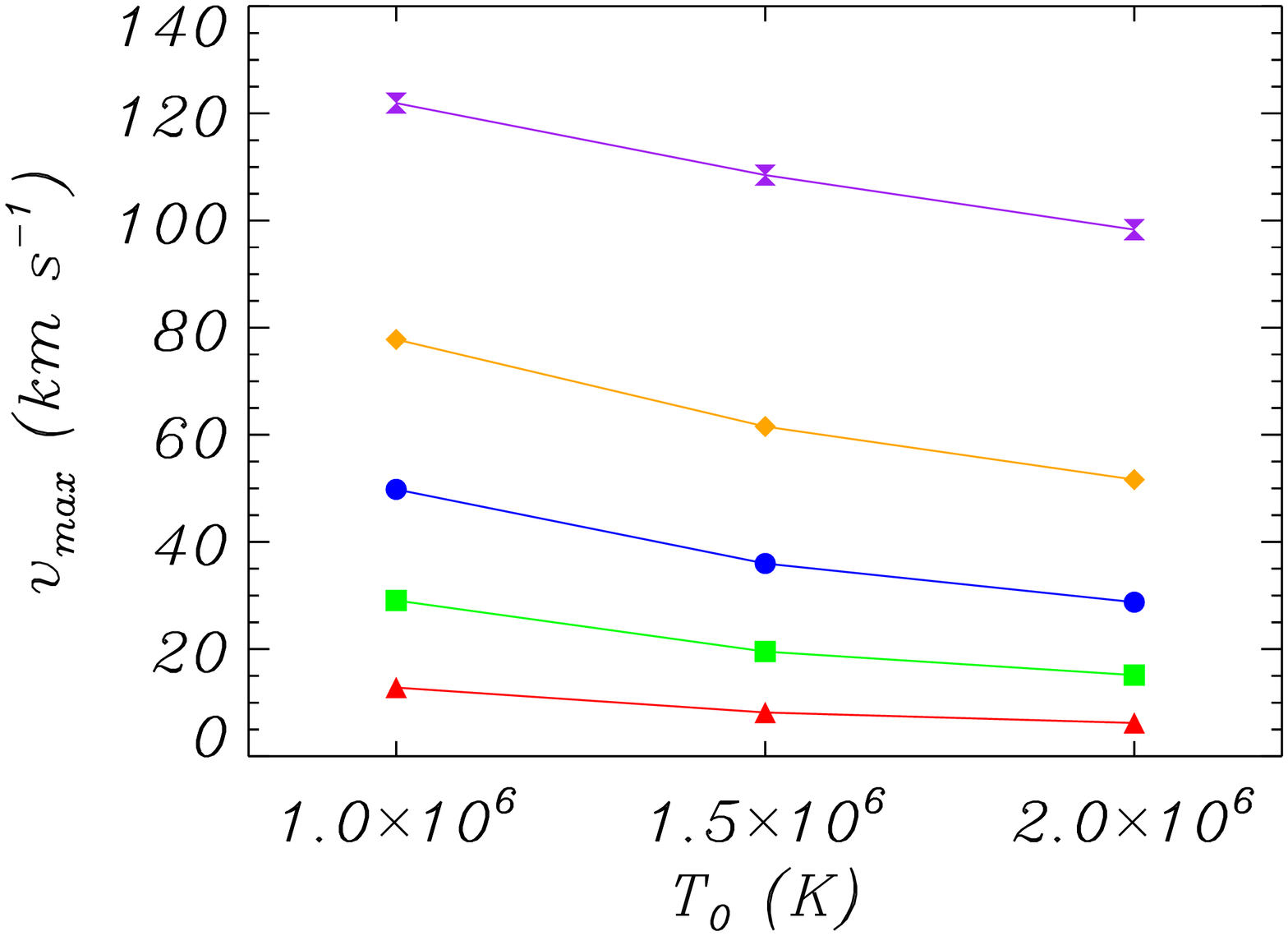}
  }
  \caption{Maximum blob velocity (in absolute value) (a) as a function of the initial blob density for different coronal temperatures and (b) as a function of the coronal temperature for different initial blob densities. From bottom to top the curves correspond to $\rhobiO=0.4,1,2,4,10\times 10^{-10}$~kg~m$^{-3}$.}
  \label{fip_velocity}
\end{figure}

Thus, the results presented in this section show that the blob dynamics is governed both by the density contrast of the blob with respect to the ambient plasma and by its mass. Since the observations of cold descending condensations provide both the blob height and speed versus time, here we also plot the maximum (i.e., most negative) descending velocity, $\vmax$. Figure~\ref{fip_velocity}a shows that the strongest variation of the maximum speed occurs for the smaller values of $\rhobiO$ in this plot, whereas the rate of change of $\vmax$ is smaller on the right side of the plot. This confirms the prediction we made before. Figure~\ref{fip_velocity}a emphasises the relevance of the blob density (or the density ratio) in the blob dynamics. On the other hand (Figure~\ref{fip_velocity}b), changing the temperature in the range $1-2\times 10^6$~K has not such a big influence, although for small values of the maximum blob density it can cause $\vmax$ to vary by a factor 1.5--2.

\section{Path curvature}
\label{sect_fip_curvature}

In this section we are concerned with a fully ionised plasma blob falling along circular magnetic field lines of radius $R$. We assume a very large electric conductivity, thus ensuring that the plasma cannot move across magnetic field lines. In addition, we assume a low-$\beta$ situation so that the plasma cannot drag the magnetic field with it. Instead, the magnetic field maintains its shape and guides plasma motions.

\begin{figure}[h]
  \centerline{
    \includegraphics[width=0.33\textwidth,angle=-90]{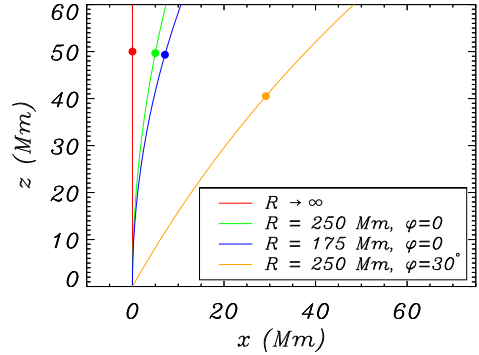}
  }
  \caption{Four trajectories of descending plasma condensations. The red, green, and blue lines correspond to circular paths of different curvature that become vertical at the coronal base, whereas the orange curve is also circular but makes a $30^{\rm o}$ angle with the vertical direction at $z=0$. The dot on each curve marks the initial blob position in the numerical simulations. Line colors correspond to those used in Figure~\ref{fip_curvature}.}
  \label{curved_path}
\end{figure}

We impose that magnetic field lines make an angle $\varphi$ with the vertical direction at $z=0$ and so the height of a point in the falling trajectory is given by

\begin{equation}\label{eqxzcircular}
z=R\left[\sin\left(\frac{s}{R}+\varphi\right)-\sin\varphi\right],
\end{equation}

\noindent with $s$ the distance from the $z=0$ level to this point along the field line. The paths used in this section are shown in Figure~\ref{curved_path}.

We next use curvilinear coordinates associated to magnetic field lines and consider Equations~(\ref{eqmassi}) and (\ref{eqpressie}) together with the component of Equation~(\ref{eqmomi}) parallel to the circular path. We obtain

\begin{equation}\label{eqrhois}
\frac{\partial\rhoi}{\partial t}=-\vi\frac{\partial\rhoi}{\partial s}-\rhoi\frac{\partial\vi}{\partial s},
\end{equation}

\begin{equation}\label{eqvis}
\rhoi\frac{\partial\vi}{\partial t}=-\rhoi\vi\frac{\partial\vi}{\partial s}-\frac{\partial\pie}{\partial s}-\gs\rhoi,
\end{equation}

\begin{equation}\label{eqpies}
\frac{\partial\pie}{\partial t}=-\vi\frac{\partial\pie}{\partial s}-\gamma\pie\frac{\partial\vi}{\partial s},
\end{equation}

\noindent were the unknowns $\rhoi$, $\vi$, and $\pie$ here depend on $s$ and $t$ and $\gs=g\cos(s/R+\varphi)$ is the component of the acceleration of gravity parallel to the circular path. Note that in Equations~(\ref{eqrhoi}) to (\ref{eqpie}) $\vi$ represents the vertical velocity, but in Equations~(\ref{eqrhois}) to (\ref{eqpies}) it represents the velocity component parallel to the curved path.

The plasma configuration at $t=0$ is obtained by setting $\vi=0$ and ignoring time variations. Then, Equations~(\ref{eqrhois}) and (\ref{eqpies}) are automatically satisfied and Equation~(\ref{eqvis}) reduces to

\begin{equation}\label{equilins}
0=-\frac{\partial\pie}{\partial s}-\gs\rhoi.
\end{equation}

\noindent Using the ideal gas law for charged particles, their pressure and density distribution along the circular trajectory at $t=0$ is

\begin{equation}\label{sol_equilis}
\pie(s,t=0)=\pieO\exp\left[-\frac{z(s)}{\Hi}\right], \hspace{0.3cm} \rhoi(s,t=0)=\rhoiO\exp\left[-\frac{z(s)}{\Hi}\right],
\end{equation}

\noindent where the vertical scale height, $\Hi$, has the same value as for the vertical trajectory and $z(s)$ is given by Equation~(\ref{eqxzcircular}). This formula is formally identical to its counterpart for a vertical path (Equation~(\ref{sol_equili})).

Regarding the mass condensation at $t=0$, we use the following expression (which is analogous to Equation~(\ref{rhoblob}))

\begin{equation}\label{rhoblobis}
\rhobi(s,t=0)=\rhobiO\exp\left[-\left(\frac{s-\sO}{\Delta}\right)^2\right].
\end{equation}

\noindent Hence, the initial distance of the blob center from $z=0$ along the curved magnetic field lines is $\sO$, although its height is obviously smaller than this value; cf. Equation~(\ref{eqxzcircular}).

This initial plasma configuration (vertically stratified coronal atmosphere and embedded mass condensation) is identical to the one we have considered so far if one substitutes $s$ for $z$ and $\sO$ for $\zO$. Equations~(\ref{eqrhois}) to (\ref{eqpies}) are also formally identical to their counterparts for vertical motions (i.e., Equations~(\ref{eqrhoi}), (\ref{eqvi}), and (\ref{eqpie})), except for the projected acceleration of gravity along the circular magnetic field lines, $\gs$. We can thus expect the effect of curvature to be noticeable only if the path is inclined from the vertical direction, so that $\gs$ departs significantly from $g$. Moreover, if the condensation travels a long distance along the magnetic field then the effective acceleration of gravity will increase during this motion and this will also influence its dynamics.

\begin{figure}[t!]
  \centerline{
    \scriptsize{(a)}
    \includegraphics[width=0.33\textwidth,angle=-90]{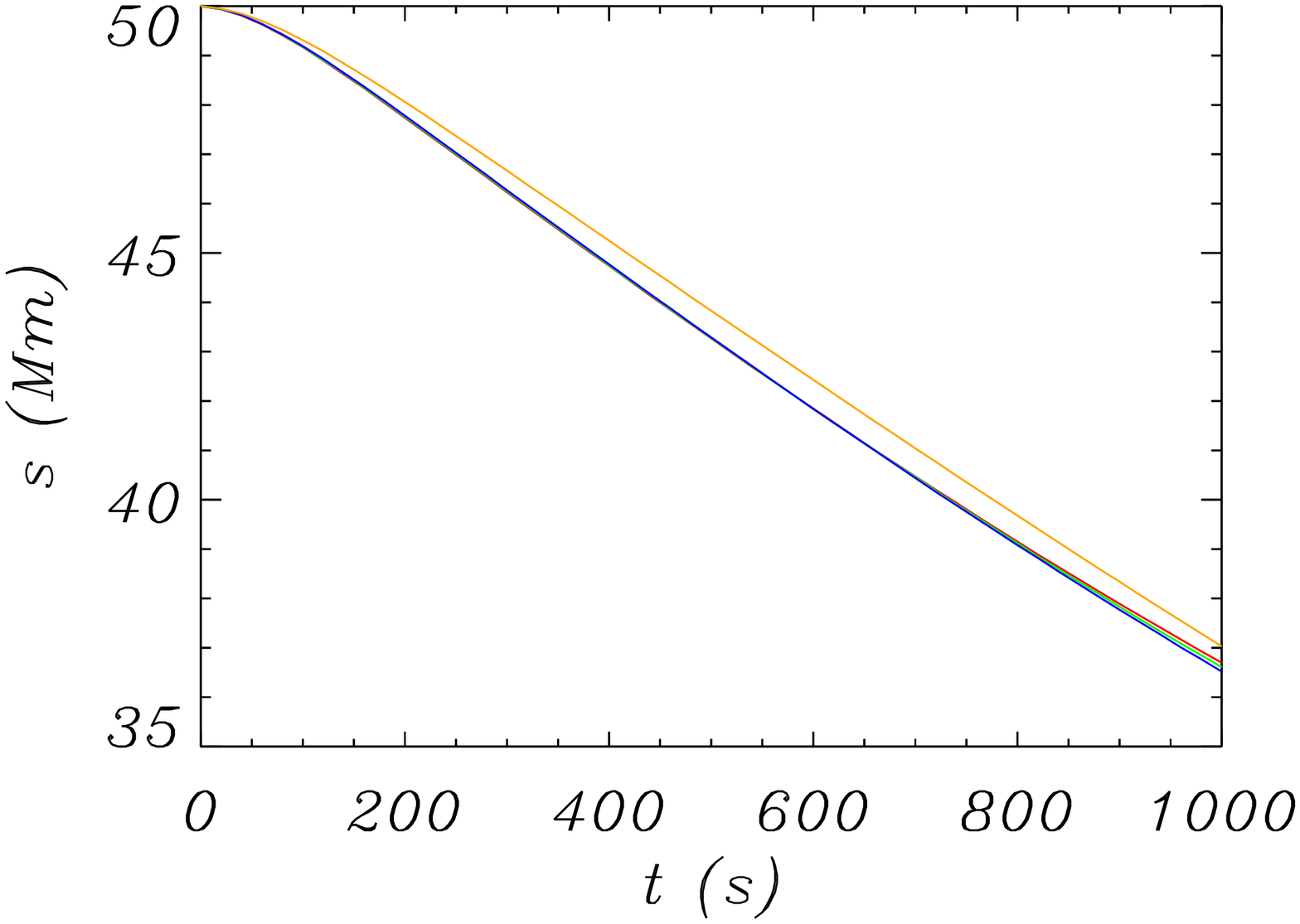}
    \scriptsize{(b)}
    \includegraphics[width=0.33\textwidth,angle=-90]{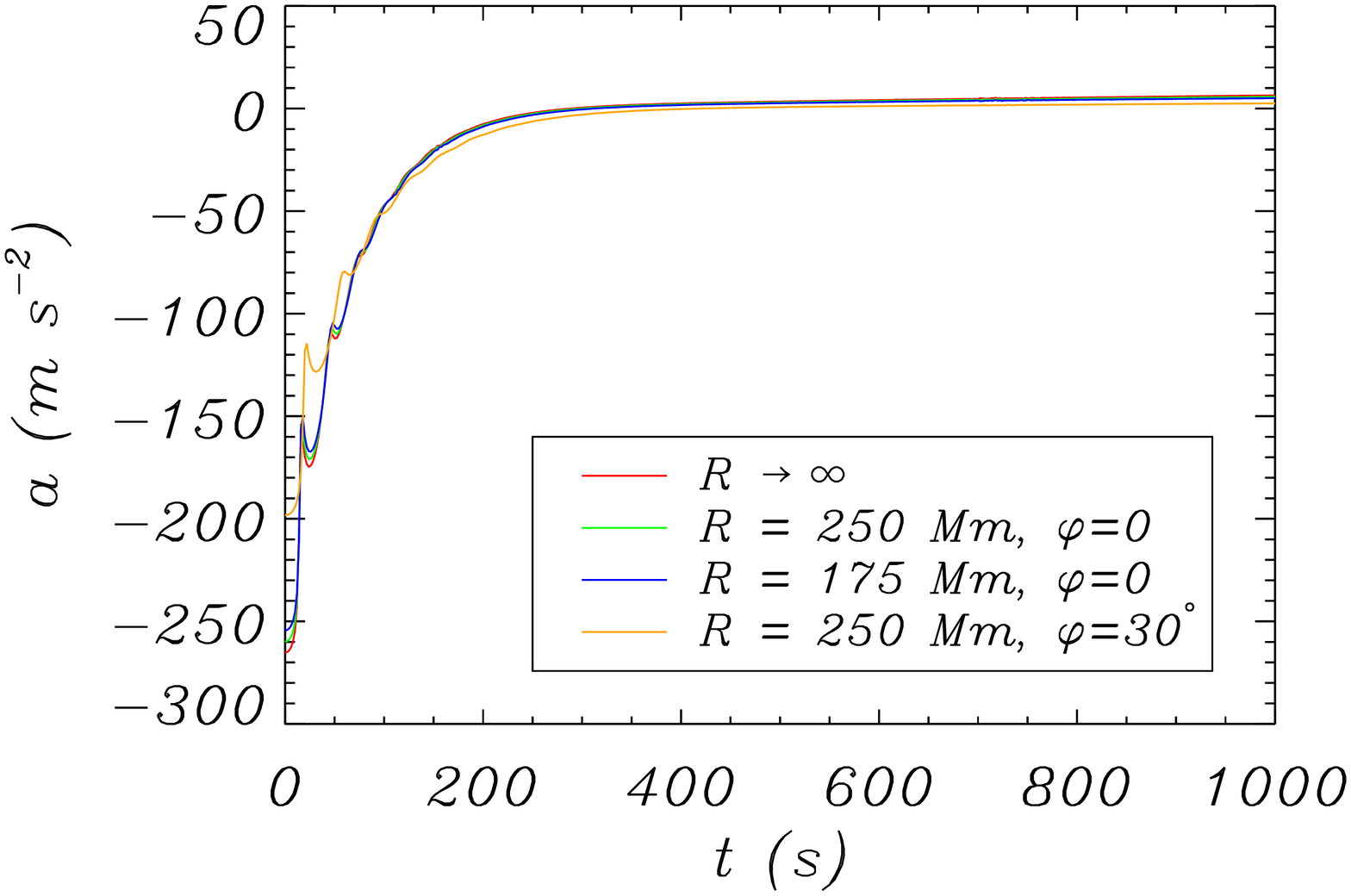}
  }
  \centerline{
    \scriptsize{(c)}
    \includegraphics[width=0.33\textwidth,angle=-90]{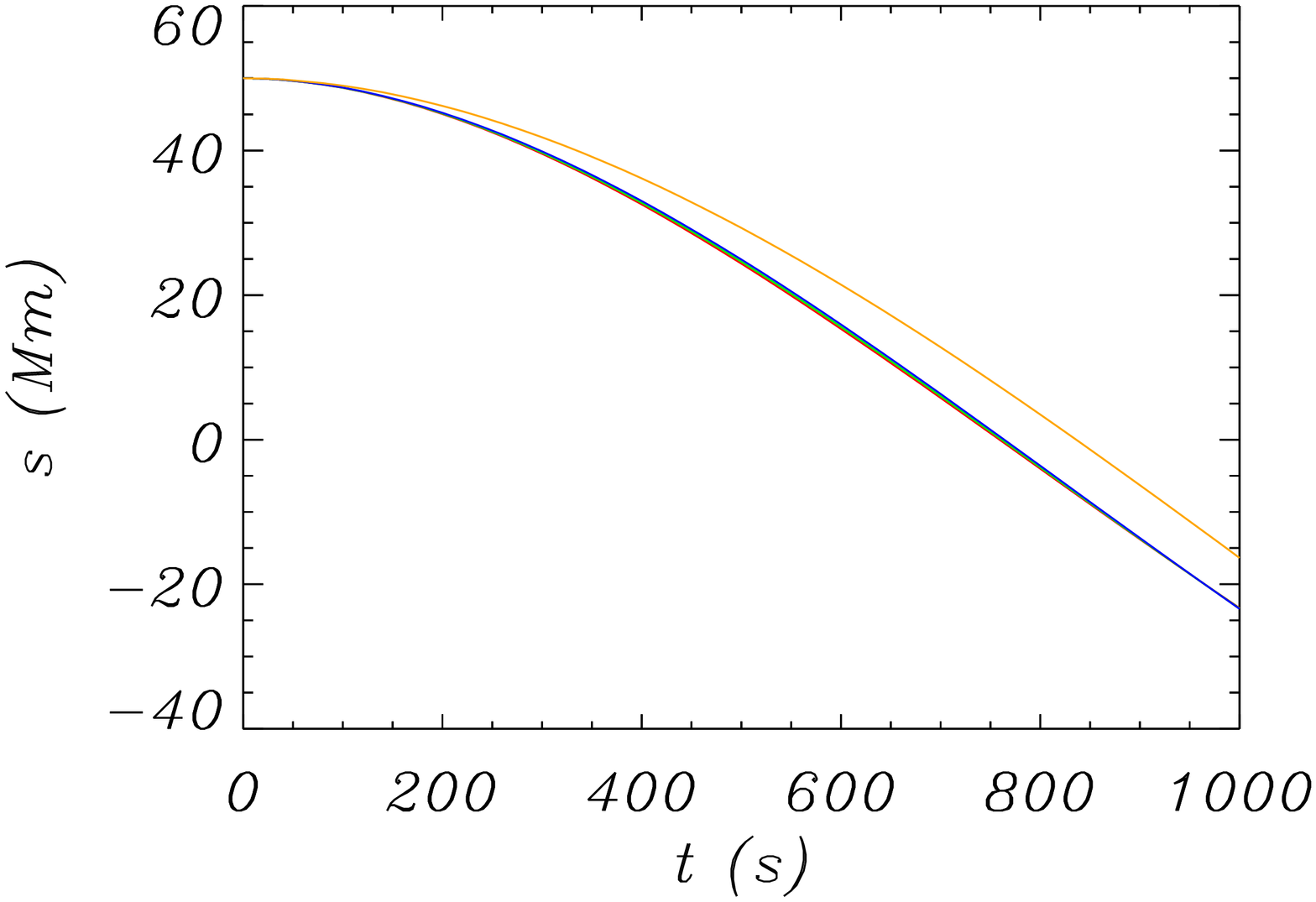}
    \scriptsize{(d)}
    \includegraphics[width=0.33\textwidth,angle=-90]{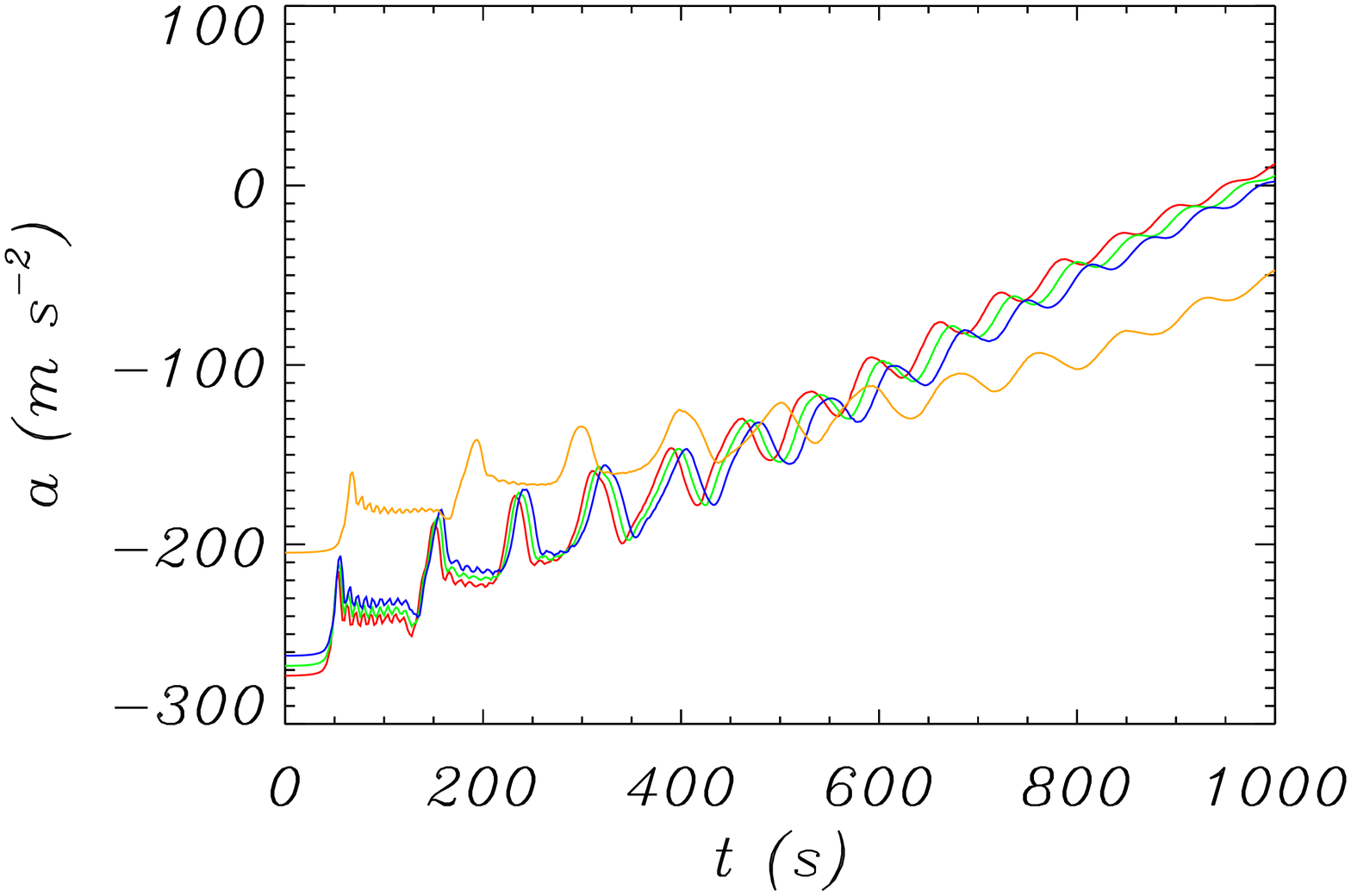}
  }
  \caption{(a) and (c) Blob position and (b) and (d) blob acceleration along the descending trajectory as a function of time. The red lines correspond to a vertical falling motion (i.e., to the results presented in \S~\ref{sect_fip_blob_dynamics} and \ref{sect_fip_blob_dynamics10}) and so $s$ is equal to the blob height, while the other curves correspond to motions along a curved path of radius $R$ and angle $\varphi$ with the vertical direction at $z=0$. In this case, $s$ is the blob distance from $z=0$ along the curved path. (a) and (b) correspond to the initial maximum blob density $\rhobiO=10^{-10}$~kg~m$^{-3}$; (c) and (d) are for $\rhobiO=10^{-9}$~kg~m$^{-3}$. The meaning of all line colors is shown in panel (b).}
  \label{fip_curvature}
\end{figure}

We have solved Equations~(\ref{eqrhois}) to (\ref{eqpies}) using the parameter values of \S~\ref{sect_fip_blob_dynamics} and \ref{sect_fip_blob_dynamics10} and for two different radii of magnetic field lines ($R=175$~Mm and 250~Mm) and two different inclinations of magnetic field lines at $z=0$ (namely $\varphi=0$ and $\varphi=30^{\rm o}$). To explore the blob dynamics, in Figure~\ref{fip_curvature}a we represent the blob height (for a vertical path) or its distance from the $z=0$ level along magnetic field lines (for a circular trajectory). The time-distance variation of the four cases is rather similar and, except for the one with inclined field lines at the coronal base, the trajectories of the other three are almost undistinguishable. A small departure can only be appreciated at the end of the numerical simulation. The variation of the blob acceleration with time, Figure~\ref{fip_curvature}b, displays an initial value of $a$ equal to the projection of $g$ on field lines. For the case $R=250$~Mm, $\varphi=30^{\rm o}$ this initial acceleration is of the order of 200~m~s$^{-2}$, whereas for the curved  field lines that are vertical at the coronal base it is much closer to the vertical acceleration of gravity. In spite of these differences in the starting value of $a$, in a very short time ($\simeq 25$~s and $\simeq 100$~s for the cases $\varphi=0$ and $\varphi=30^{\rm o}$, respectively) the blob moving along a curved trajectory displays an acceleration that closely matches that of the vertically falling condensation.

These results confirm our previous speculations about the effect of curvature. We hypothesised that the path inclination would have some effect only if $\gs$ is different enough from $g$. The two cases in Figures~\ref{fip_curvature}a and b that coincide with the vertical motion, i.e., the curved lines with $R=175$~Mm and 250~Mm, $\varphi=0$, have $\gs$ that differs 4\% from $g$ at most along the blob track. On the other hand, the case with $\varphi=30^{\rm o}$ displays a maximum difference between $\gs$ and $g$ of 25\%. It must be emphasised, nevertheless, that the interplay between the pressure gradient and gravity leads in the four cases to a similar blob acceleration after 100~s at most.

To complete our study of the effect of the path curvature we make use of the higher blob density considered above, $\rhobiO=10^{-9}$~kg~m$^{-3}$. The variation of blob position and the acceleration along the descending path versus time are shown in Figures~\ref{fip_curvature}c and d. Just as happens with the case $\rhobiO=10^{-10}$~kg~m$^{-3}$ only the blob falling along the curved magnetic field with an angle $\varphi=30^{\rm o}$ at $z=0$ descends with a different dynamics, although its departure from the other three curves is much larger in Figure~\ref{fip_curvature}c than in Figure~\ref{fip_curvature}a. Moreover, the acceleration for $\varphi=30^{\rm o}$ is far from that of the other three cases because having a larger density ratio implies that the pressure structure takes much longer to readjust in order for its gradient to compensate the gravity force. An additional effect that appears for $\varphi=30^{\rm o}$ is that the inclination of field lines changes during the blob motion so as to increase $\gs$ and this requires a continuous readjustment of the pressure configuration in order to reach the balance with gravity.

\section{Conclusions}
\label{sect_conclusions}

In this work we have investigated some physical processes present in dense downfalling plasma blobs such as coronal rain and dense prominence knots. To simplify our task several important ingredients have been ignored: the formation process of the blobs, the presence of neutrals (that will be included in the future), non-ideal effects, and the effect of magnetic fields (apart from behaving as rigid guides for plasma motions). Moreover, the influence of the chromosphere and photosphere has also been omitted by putting the numerical boundaries far from the range of heights of interest. Our model thus contemplates the dynamical evolution of a fully ionised plasma condensation in an isothermal, vertically stratified corona, with motions that are either vertical or guided along rigid magnetic field lines (with assumed circular shape).

We have first focused our attention on a blob falling vertically. The mass condensation is released at rest and starts to accelerate, showing two distinct phases. In the first one, gravity is the dominant force and causes the downward acceleration of the blob. At $t=0$ a sound wave is emitted at the initial blob position. This wave results in a considerable modification of the coronal pressure, by which a strong pressure gradient is generated at the blob position. At the end of the first phase this gradient, that moves with the mass condensation as it falls down, becomes large enough to cancel the acceleration of gravity and the blob enters the second phase, characterised by a roughly constant velocity. This two-stage dynamics is very often found in time-distance diagrams of coronal rain events, as for example Figure~6 of \citet{degroof04}, Figure~5 of \citet{degroof05}, Figure~3 of \citet{antolin10b}, and Figure~7 of \citet{antolin12a}. Moreover, \citet{schrijver01} has represented the temporal variation of the descending speed for 45 coronal rain features and many of them clearly display the initial acceleration phase. This remarkable similarity between our numerical simulations and observations points out that the simulations take into account some of the relevant physics for the dynamics of descending blobs.

A larger blob mass requires a larger pressure gradient to balance gravity and this leads to a lengthening of the downward acceleration phase, that can last between 100~s and 1000~s. A longer acceleration phase in turn allows the blob to achieve higher velocity and so we have found an increase of the maximum falling speed with the initial blob density ($\rhobiO$). For the parameter values used in this work, this maximum speed is more sensitive to changes of $\rhobiO$ around $10^{-10}$~kg~m$^{-3}$ than to changes around $10^{-9}$~kg~m$^{-3}$. Another important conclusion refers to the density contrast, that for a fixed coronal temperature is determined by the initial blob density and the density at the base of the corona. If these two parameters are modified simultaneously in such a way that the density contrast is held fixed, then the blob dynamics does not change. And finally, we have also established that the coronal temperature also has its influence because it determines the coronal vertical scale height. We have seen that a colder corona has a smaller vertical scale height and so, for a given density at $z=0$, the mass condensation moves in a rarer environment compared to that of a hotter corona. Hence, this situation is similar to that of a blob with a higher density contrast, so that the condensation descends faster in a cold corona than in a hotter one.

Blob speeds in the range 10--150~km~s$^{-1}$ have been obtained in this work. This velocity range covers all observations of coronal rain events and falling prominence knots, but a detailed comparison with observations is not pertinent because, as discussed in the previous paragraph, a given blob velocity can be obtained with different combinations of our parameters. Nevertheless, other results in our simulations could be compared to observations. For example, although the blob mass is conserved in time, its maximum density has been found to build up. In particular, for a plasma condensation with density of the order of $10^{-10}$~kg~m$^{-3}$ this increase is about 6\% after 1000~s, while for larger initial blob densities (of the order of $10^{-9}$~kg~m$^{-3}$) it is above 60\%. In this last case the condensation density also presents a strong distortion with respect to its initial profile: the density variation at the blob bottom becomes steeper with time and so a large enough gradient could develop such that a shock might appear. Another feature of descending condensations is that they emit small amplitude sound waves as they fall. This wave emission is completely different from the one observed at $t=0$, that distorts the background corona, and is present during the whole simulation. We have performed a simple perturbation analysis of this phenomenon and have derived leaky wave periods that are in good agreement with those found in the simulations. The sound wave emission during the blob lifetime is more evident for larger blob densities and might be observed as small periodic changes of the blob intensity. A trend subtraction could be necessary to achieve this detection.

Coronal rain blobs are observed to move along curved trajectories with a variety of shapes and inclinations \citep[see e.g.,][]{schrijver01,antolin12a}. It has been found that loop curvature is irrelevant for the blob dynamics if magnetic field lines are vertical at the coronal base. Noticeable differences are only present if magnetic field lines make an angle with the vertical direction at $z=0$ (here the value $\varphi=30^{\rm o}$ has been used) and if the condensation to corona density ratio is large enough. In view of these results, one could argue that the coronal rain condensations detected by \citet{schrijver01}, whose acceleration depends very little on the path inclination, had a moderate density and/or small path inclination. It is of great interest to solve the two-dimensional problem of partially ionised mass condensations falling along curved magnetic fields to see if the coupling between ions and neutrals is strong enough to drag neutrals along the magnetic field and to prevent them from falling vertically. In fact, observations seem to indicate that this must be the case since coronal rain blobs detected in the H$\alpha$ line usually display curved trajectories. For a numerical study of fully ionised plasma blobs moving in a magnetic arcade the reader is referred to \citet{fang13}.

The motion of individual plasma elements has also been investigated and it has been found that the position of blob maximum density moves faster than nearby particles. Thus, the background material penetrates inside the blob from below while the blob material can leave it at its top and become part of the background plasma. This is more than just a curiosity. In a partially ionised plasma it may lead to the recombination (ionisation) of material entering (leaving) the blob and so this may add more complexity to the interaction between neutrals and charged particles. This will be investigated in a  forthcoming article.

The blob temperature remains constant during the whole simulation regardless of its initial density. It is worth mentioning that our model does not include thermal conduction, that could smear out the large temperature difference between the plasma condensation and the environment. However, radiative losses are very efficient at low temperatures and not only lead to the blob formation by catastrophic cooling, but they also help the condensation to keep its temperature after being formed \citep{muller04,muller05,antolin10b,fang13}.

We finish with a comment on our one-dimensional assumption. If motions are guided by vertical or curved magnetic fields, then this assumption is reasonable because the material moving inside a magnetic tube is decoupled from that outside it. Nevertheless, in the absence of a magnetic field the blob behaves like a piston moving in a liquid. The piston presses the liquid, that can move around the piston, so that the strong pressure gradient we found in our simulations could actually be substantially smaller and blob speeds substantially larger. The second phase of blob dynamics, namely that with roughly constant speed, would also be modified or could even disappear. This is a motivation for undertaking two- or three-dimensional simulations.

\begin{acknowledgements}
RO, RS, and JT acknowledge support from MINECO and FEDER funds through project AYA2011-22846 and from CAIB through the ``Grups Competitius'' program and FEDER funds. The work of TZ was also supported by FP7-PEOPLE-2010-IRSES-269299 project SOLSPANET and by Shota Rustaveli National Science Foundation grant DI/14/6-310/12. MK acknowledges the projects P25587-N27 and S11606-N16 of the Austrian Science Foundation (FWF). JT acknowledges support from MINECO through a Ram\'on y Cajal grant. Wavelet software was provided by C. Torrence and G. Compo, and is available at http://atoc.colorado.edu/research/wavelets/. RO also thanks D. W. Fanning for making available the Coyote Library of IDL programs (http://www.idlcoyote.com/).
\end{acknowledgements}

\appendix

\section{Small amplitude leaky sound waves in a dense blob}
\label{sect_appendix}

The usual procedure to study small amplitude trapped and leaky waves is to start from an equilibrium state and next to consider linear perturbations about this equilibrium. The problem here is that we want to study waves emitted by the blob, that is not in an equilibrium. For this reason, our results can only provide an approximation to the actual wave leakage found in the numerical simulations.

We neglect gravity and consider a static equilibrium with uniform pressure ($\pO$) and a density consisting of a uniform background ($\rhoe$) plus a condensation of the form of Equation~(\ref{rhoblob}). This equilibrium state is defined by

\begin{equation}\label{wave_equil}
\pie(z)=\pO, \hspace{1cm} \rhoi(z)\equiv\rhoO(z)=\rhoe+\rhobiO\exp\left[-\left(\frac{z}{\Delta}\right)^2\right], \hspace{1cm} \vi(z)=0,
\end{equation}

\noindent where $z=0$ now gives the central, unperturbed blob position. The next step is to take perturbations about this equilibrium and, given that we are interested in periodic variations, their time dependence is of the form $\exp(i\omega t)$. The pressure, density, and velocity perturbations are respectively given by the functions $\pU(z)$, $\rhoU(z)$, and $\vU(z)$ multiplied by $\exp(i\omega t)$. With these assumptions, the linearised form of Equations~(\ref{eqrhoi})--(\ref{eqpie}) is

\begin{equation}\label{wave_eq_rho1}
i\omega\rhoU=-\frac{d\rhoO}{dz}\vU-\rhoO\frac{d\vU}{dz},
\end{equation}

\begin{equation}\label{wave_eq_v1}
i\omega\rhoO\vU=-\frac{d\pU}{dz},
\end{equation}

\begin{equation}\label{wave_eq_p1}
i\omega\pU=-\gamma\pO\frac{d\vU}{dz}.
\end{equation}

\noindent Equations~(\ref{wave_eq_v1}) and (\ref{wave_eq_p1}) can be combined to give

\begin{equation}\label{wave_eq_v1-2}
\frac{d^2\vU}{dz^2}+\frac{\omega^2}{\cs^2}\vU=0,
\end{equation}

\noindent where the sound speed squared is defined as $\cs^2=\gamma\pO/\rhoO$. The sound speed has a minimum value $\gamma\pO/(\rhoe+\rhobiO)$ at the blob center and achieves its maximum value $\gamma\pO/\rhoe$ as $|z|\rightarrow\infty$. These two values of $\cs$ are denoted by $\csi$ and $\cse$, respectively.

The two possible types of solutions to Equation~(\ref{wave_eq_v1-2}) are trapped waves, for which $\vU\rightarrow 0$ as $|z|\rightarrow\infty$ and $\omega$ is real, and leaky waves, for which $\vU$ oscillates and grows as $|z|\rightarrow\infty$ and $\omega$ is complex with positive imaginary part. Moreover, both trapped and leaky solutions can either have $\vU$ even and $\rhoU$, $\pU$ odd about the blob center or $\vU$ odd and $\rhoU$, $\pU$ even about $z=0$. We refer to these solutions simply as even and odd solutions.

Solutions to Equation~(\ref{wave_eq_v1-2}) are obtained numerically using the well-known shooting and matching technique, with the boundary conditions obtained as follows. At a large distance from the plasma condensation the medium can be considered uniform and with a sound speed equal to $\cse$. Then, the plasma velocity is $\vU(z)=A_+\exp(i\omega z/\cse)+A_-\exp(-i\omega z/\cse)$. This solution represents the superposition of two waves propagating in opposite directions. If $\omega$ is real this solution is oscillatory and so the present equilibrium configuration does not support trapped sound waves. If $\omega$ is complex Equation~(\ref{wave_eq_v1-2}) has leaky solutions, characterised by a temporal attenuation and a spatial growth at large distances from the condensation. To find leaky solutions we write $\omega=\omegar+i\omegai$, where both $\omegar$ and $\omegai$ are real. In order to have perturbations that damp in time, we require $\omegai$ to be positive. We also take $\omegar>0$ with no loss of generality. Then, considering $z>0$ in the former expression for $\vU(z)$, the term with amplitude $A_+$ ($A_-$) corresponds to a wave travelling towards the blob (away from the blob), whereas the opposite applies for $z<0$. To study wave emission by the blob we must only retain waves emanating from it and so we have $\vU(z)=A_+\exp(i\omega z/\cse)$ for $z<0$ and $\vU(z)=A_-\exp(-i\omega z/\cse)$ for $z>0$. We now proceed as follows: at a large distance from the blob ($z=-10$~Mm, say) we impose that the plasma velocity and its $z$-derivative can be computed from $\vU(z)=\exp(i\omega z/\cse)$, where a guess value of $\omega$ must be provided. Note that since we deal with a linear problem, we are allowed to take $A_+=1$. Starting from these initial conditions, Equation~(\ref{wave_eq_v1-2}) is integrated to the position $z=10$~Mm, which is symmetrically placed about $z=0$ with respect to the starting point of the numerical integration. If the guess value of the frequency were correct, then we would obtain the right values of $\vU$ and $d\vU/dz$. For even solutions they can be computed from $\vU(z)=\exp(-i\omega z/\cse)$ (i.e., $A_-=A_+$), whereas for odd solutions they come from $\vU(z)=-\exp(-i\omega z/\cse)$ (i.e., $A_-=-A_+$). Nevertheless, our initial value of $\omega$ will not be correct in general, so there will be a discrepancy between the numerical and analytical solutions at $z=10$~Mm. An iterative method is then applied by which $\omegar$ and $\omegai$ are varied so that this discrepancy is reduced until convergence of the numerical method is achieved. At this point we have acceptable approximations to the frequency and $\vU(z)$.

\begin{deluxetable}{ccccccccc}
\tablecolumns{9}
\tablewidth{0pc}
\tablecaption{Real and imaginary parts of the frequency, period, and exponential damping time of the lowest frequency leaky waves of a dense blob}
\tablehead{
\multicolumn{4}{c}{Odd solutions} &   \colhead{}   &
\multicolumn{4}{c}{Even solutions} \\
\cline{1-4} \cline{6-9} \\
\colhead{$\omegar$ ($\times 10^{-3}$ s$^{-1}$)} & \colhead{$\omegai$ ($\times 10^{-3}$ s$^{-1}$)} & \colhead{$P$ (s)} & \colhead{$\taud$ (s)} &
\colhead{} &
\colhead{$\omegar$ ($\times 10^{-3}$ s$^{-1}$)} & \colhead{$\omegai$ ($\times 10^{-3}$ s$^{-1}$)} & \colhead{$P$ (s)} & \colhead{$\taud$ (s)}
}
\startdata
44.04 & 2.322 & 142.7 & 430.7 & &
79.05  & 3.436  & 79.48  & 291.1 \\
113.4  & 4.483  & 55.42  & 223.1 & &
147.4  & 5.490  & 42.62  & 182.2 \\
181.4  & 6.470  & 34.65  & 154.6 & &
215.2  & 7.430  & 29.20  & 134.6 \\
\enddata
\label{table_leake}
\end{deluxetable}

To make a comparison with the simulation of \S~\ref{sect_fip_blob_dynamics10} we take the equilibrium density and pressure equal to those of the numerical simulation at the initial blob position, i.e., $\pO=0.109$~Pa, $\rhoe=3.30\times 10^{-12}$~kg~m$^{-3}$, and $\rhobiO=10^{-9}$~kg~m$^{-3}$. Then, $\csi=13.5$~km~s$^{-1}$ and $\cse=235$~km~s$^{-1}$. The results are summarised in Table~\ref{table_leake}, that contains $\omegar$, $\omegai$, the period ($P=2\pi/\omegar$), and the exponential damping time ($\taud=1/\omegai$) of the first three even and odd leaky waves. The period of the even solutions are shown in Figure~\ref{fip_wavelet} as horizontal lines and are used in \S~\ref{sect_leaky_waves} to determine that the oscillations detected in Figure~\ref{fip_dynamics10} are caused by the emission of leaky sound waves during the blob descent.


\end{document}